\documentclass[aps,prd,twocolumn,groupedaddress,showpacs,floatfix]{revtex4}

\usepackage{graphicx}
\usepackage{dcolumn}

\def\slash#1{\mkern-1.5mu\raise0.4pt\hbox{$\not$}\mkern1.2mu #1\mkern 0.7mu}

\begin{document}

\title{Perturbative chiral violations for domain-wall QCD 
        \\ with improved gauge actions}

\author{Stefano Capitani}
\email[]{stefano.capitani@uni-graz.at}
\affiliation{Institut f\"ur Physik, FB Theoretische Physik \\
Universit\"at Graz, A-8010 Graz, Austria}

\begin{abstract}
We investigate, in the framework of perturbation theory at finite $N_s$, the 
effectiveness of improved gauge actions in suppressing the chiral violations 
of domain-wall fermions. Our calculations show substantial reductions of the 
residual mass when it is compared at the same value of the gauge coupling, 
the largest suppression being obtained when the DBW2 action is used. 
Similar effects can also be observed for a power-divergent mixing 
coefficient which is chirally suppressed. No significant reduction instead 
can be seen in the case of the difference between the vector and axial-vector 
renormalization constants when improved gauge actions are used in place of 
the plaquette action. We also find that one-loop perturbation theory is not
an adequate tool to carry out comparisons at the same energy scale (of about 
2 GeV), and in fact in this case even an enhancement of the chiral violations 
is frequently obtained.
\end{abstract}

\pacs{12.38.Gc,11.30.Rd,11.30.Qc,11.10.Gh}

\maketitle

\section{Introduction}

It has been known for some time that using in the gauge sector improved actions
instead of the simpler plaquette action can significantly reduce the amount of 
chiral symmetry breaking in Monte Carlo simulations of domain-wall fermions 
\cite{Orginos:2001xa,Aoki:2002vt} (for the latest results obtained using this
kind of actions see Refs. \cite{Aoki:2005ga,Berruto:2005hg,Aoki:2006ib,Boyle:2006pw,Lin:2006kg,Lin:2006vc,Dawson:2006qc,Orginos:2006xyzt}, 
and references therein). This effect has been related to the fact that, for a 
determined value of the lattice spacing $a$, the corresponding values of 
$\beta' = 6 (1-8 c_1)/g_0^2$ (where $c_1$ is a parameter introduced in Section
\ref{sec:pt} which describes the various actions) are larger for improved gauge
actions than for the plaquette action, and the gauge fields are correspondingly
smoother. 

Renormalization factors for domain-wall fermions with improved gauge actions 
have been calculated in one-loop perturbation theory using the asymptotic 
propagators at large $N_s$ (where $N_s$ denotes the number of points in the 
fifth dimension). The renormalization factors of the quark wave function, 
the quark mass, the bilinear quark operators and many three- and four-quark 
operators can be found in \cite{Aoki:2002iq,Nakamura:2006zx} for many improved
gauge actions, and the renormalization constant of the first moment of the 
unpolarized parton distribution has been calculated using the Iwasaki action 
in \cite{Boyle:2006pw}. Here we present the first calculations made with the 
exact theory at finite $N_s$ for this kind of gauge actions, and we consider 
some quantities which can describe chiral violations.

In this article we investigate whether perturbation theory can, at least 
qualitatively, reproduce the suppressions of the chiral violations which have 
been observed in numerical simulations when improved gauge actions are used.
In a previous perturbative work \cite{Capitani:2006kw} we have investigated 
the residual mass (together with the difference between the vector and 
axial-vector renormalization constants and with a chirally suppressed mixing 
coefficient for a deep-inelastic operator) using, for finite values of $N_s$, 
the plaquette gauge action.
Here we repeat those calculations using Symanzik improved gauge actions like
the L\"uscher-Weisz action \cite{Luscher:1984xn} and renormalization group
improved gauge actions like the Iwasaki action \cite{Iwasaki:1983ck}
and the DBW2 action \cite{Takaishi:1996xj,deForcrand:1999bi}.
By repeating the computations of the same above mentioned quantities with 
all these improved actions we can then make a direct comparison of the 
corresponding results and investigate the reduction of the chiral violations 
attained by each of these actions. As was the case for the plaquette, in 
carrying out these investigations one of our principal aims is also to 
calculate how these quantities behave for several choices of $N_s$ and of the 
domain-wall height (or Dirac mass) $M$. 

Apart from the new gluon propagators, the basic features of the calculations 
presented here remain the same as in \cite{Capitani:2006kw}, and in order not 
to render this paper too cumbersome we do not show here again many of the 
Tables and formulae that are independent of the gauge action used and that 
can be found in that paper. This applies in particular to the $1-\lambda$ 
contributions (in a general covariant gauge described by $\lambda$) at one 
loop, and to the anomalous dimensions. 
For notations, the action and the basic fermion propagators we also refer to 
\cite{Capitani:2006kw}, which the reader is invited to consult for a fuller 
understanding of the calculations carried out in the present paper.

This article is organized as follows. In Sect. \ref{sec:pt} we give the
explicit expression of the gluon propagators for the class of improved gauge 
actions that we employ, and in Sect. \ref{sec:rm} we then present the results 
for the residual mass at finite $N_s$, showing how its suppression is 
achieved by these actions when the coupling is kept constant. 
In Sect. \ref{sec:vad1} we give the results for 
the difference between the vector and axial-vector renormalization constants 
and for the power-divergent mixing (due to the breaking of chiral symmetry) 
of an operator which describes polarized parton distributions. In this case
we find that not always the use of improved gauge actions produces a 
significant reduction in the magnitude of the results at the same gauge 
coupling. Finally, in Sect. 
\ref{sec:concl} we make some concluding remarks. An Appendix reports a few 
Tables concerning results for the plaquette action, so as to ease comparisons 
and make clearer the reductions of the chiral violations achieved by improved 
gauge actions compared with the plaquette action.

\section{Perturbation theory}
\label{sec:pt}

We use the standard formulation of domain-wall fermions of Shamir 
\cite{Shamir:1993zy} where the quarks are massless, 
\begin{eqnarray}
S^{DW}_q &=& \sum_x \sum_{s=1}^{N_s} \Bigg[
\frac{1}{2} \sum_\mu \Big( \overline{\psi}_s(x)
(\gamma_\mu +1) U_\mu (x) \psi_s(x+\hat{\mu})
\nonumber \\
&& - \overline{\psi}_s(x)
(\gamma_\mu -1) U^\dagger_\mu (x-\hat{\mu}) \psi_s(x-\hat{\mu}) \Big)
\nonumber \\
&& + \Big( \overline{\psi}_s(x) P_+ \psi_{s+1}(x)
         + \overline{\psi}_s(x) P_- \psi_{s-1}(x) \Big)
\nonumber \\
&& + (M -1 -4) \overline{\psi}_s(x) \psi_s(x) \Bigg] , 
\label{eq:dwaction}
\end{eqnarray}
and with improved gauge actions in place of the Wilson plaquette action in 
the gauge sector. For what concerns the Feynman rules, the fermionic part is 
then the same as in \cite{Capitani:2006kw}, including the expressions for the 
quark-gluon vertices. The gluon propagators instead are different, and in 
covariant gauge their expression in momentum space is \cite{Weisz:1983bn}
\begin{widetext}
\begin{equation}
G_{\mu\nu} (k) = \frac{1}{(\widehat{k}^2)^2} \,
\Big( (1-A_{\mu\nu} (k)) \, \widehat{k}_\mu \widehat{k}_\nu +\delta_{\mu\nu} 
   \, \sum_\sigma \, \widehat{k}^2_\sigma \, A_{\nu\sigma} (k) \Big) 
 - (1-\lambda ) \, \frac{\widehat{k}_\mu \widehat{k}_\nu}{(\widehat{k}^2)^2} ,
\end{equation}
where $\widehat{k}_\mu = 2 \sin k_\mu/2$, and $A_{\mu\nu}$ is symmetric 
in $\mu$ and $\nu$ and given by
\begin{equation}
A_{\mu\nu} (k) = \frac{1-\delta_{\mu\nu}}{\Delta (k)} \,
\Big[ (\widehat{k}^2)^2 
-c_1 \widehat{k}^2 \Big( 2\sum_\rho \widehat{k}_\rho^4
+\widehat{k}^2 \sum_{\rho \neq \mu,\nu} \widehat{k}_\rho ^2 \Big) 
+ c_1^2 \Big( \Big( \sum_\rho \widehat{k}_\rho^4 \Big)^2 
+\widehat{k}^2 \sum_\rho \widehat{k}_\rho^4 \sum_{\tau \neq \mu,\nu} 
\widehat{k}_\tau^2 + (\widehat{k}^2)^2 \prod_{\rho \neq \mu,\nu} 
\widehat{k}_\rho^2 \Big) \Big] ,
\end{equation}
with
\begin{equation}
\Delta (k) = \Big( \widehat{k}^2 -c_1 \sum_\rho \widehat{k}_\rho^4 \Big)
\Big[ \widehat{k}^2 
-c_1 \Big( (\widehat{k}^2)^2 + \sum_\tau \widehat{k}_\tau^4 \Big) 
+\frac{1}{2} c_1^2 \Big( (\widehat{k}^2)^3 + 2 \sum_\tau \widehat{k}_\tau^6 
- \widehat{k}^2 \sum_\tau \widehat{k}_\tau^4 \Big) \Big] 
-4 c_1^3 \sum_\rho \widehat{k}_\rho^4 \prod_{\tau \neq \rho} 
\widehat{k}_\tau^2 .
\end{equation}
\end{widetext}
The parameter $c_1$ describes the various actions: the choice $c_1 = -1/12$ 
corresponds to the L\"uscher-Weisz action, $c_1 = -0.331$ corresponds to the 
Iwasaki action and $c_1 = -1.40686$ corresponds to the DBW2 action. Putting 
$c_1=0$ one recovers the expression of the Wilson plaquette propagator. 
For completeness, we mention that for $c_1 \neq 0$ the gluon vertices are also 
different from the ones of the plaquette action, and in particular the 
expressions of the 3- and 4-gluon vertices can be found in \cite{Weisz:1983bn}.
However, these new gluon vertices are only needed beyond one loop for the 
quantities investigated in this paper, and thus they will not interest us here.

The above expression for the gluon propagators clearly shows that, for a 
generic one-loop matrix element between quark states, the part proportional 
to $1-\lambda$, that is the difference between its results for the Landau and 
Feynman gauges, is exactly equal to what one obtains using the plaquette 
action. For this reason we do not need to compute this part again and report
here the results referring to it. These numbers can be found in the appropriate
Tables of \cite{Capitani:2006kw}.

We find it useful to remind that in \cite{Capitani:2006kw} it was also found 
that for many quantities the part proportional to $1-\lambda$ presents a 
nontrivial dependence on $N_s$ and $M$. Thus, also in the case of improved 
gauge actions the results for the residual mass and in general for the 
renormalization factors and mixing coefficients turn out not to be gauge 
invariant. Furthermore, the anomalous dimensions of operators at finite
$N_s$ are also the same as in the plaquette case, and they are thus different 
from their continuum values and again depend on $N_s$ and $M$.

In our computations we have used the symbolic manipulation program FORM 
\cite{Vermaseren:2000nd} to perform the algebraic calculations, integrating 
afterwards the corresponding expressions by means of Fortran codes, as 
explained in \cite{Capitani:2006kw}.

\section{Residual mass}
\label{sec:rm}

Although the quark fields are originally massless in the Lagrangian that we 
employ, the truncation of domain-wall fermions at finite $N_s$ generates 
already at tree level a nonvanishing residual mass of the physical fields  
\cite{Vranas:1997da,Vranas:1997ib,Capitani:2006kw}:  
\begin{equation}
a \, m_{res}^{(0)} = - w_0^{N_s} (1-w_0^2) = - (1-M)^{N_s} \, M(2-M) .
\label{eq:mres_tree}
\end{equation}
From now on we use the abbreviation $w_0 = 1-M$, and we remind that the 
physical fields that we use are the standard ones and given by
\begin{eqnarray}
q(x) &=& P_+ \psi_1 (x) + P_- \psi_{N_s} (x) \\
\overline q(x) &=& \overline\psi_{N_s} (x) P_+ + \overline\psi_1 (x) P_- .
\end{eqnarray}
Since we work with even $N_s$, $m_{res}^{(0)}$ is always a negative quantity. 
Its values for several choices of $N_s$ and $M$ are collected in Table 
\ref{tab:residual_tree} in the Appendix, where they have already been 
multiplied for $16 \pi^2$, so as to make them more homogeneous with the numbers
that we then report for the one-loop diagrams.

Radiative corrections provide additional contributions to $m_{res}$. 
At one loop the critical (or residual) mass is determined by the formula
\begin{equation}
a \, m_{res}^{(1)} = - w_0^{N_s}(1-w_0^2) - \bar g^2 \, \Sigma_0 ,
\label{eq:mres}
\end{equation}
where $\Sigma_0$ is the term proportional to $1/a$ in the self-energy of the
quark,
\begin{eqnarray}
\Sigma_q (p) &=&  \frac{\bar g^2}{1-w_0^2} \, \Big[ \frac{\Sigma_0}{a}  
+ i\slash{p} \, \Big( c_{\Sigma_1}^{(N_s,M)} \log a^2 p^2 + \Sigma_1 \Big)
\nonumber \\ && \qquad 
- \big( i\slash{p}-w_0^{N_s}(1-w_0^2) \big) \, 
\frac{2w_0}{1-w_0^2} \, \Sigma_3 \Big] . 
\label{eq:seint}
\end{eqnarray}
Indeed, at this order one can write 
\begin{eqnarray}
\langle q (-p) \overline q (p)\rangle_{1~loop}
&=& \frac{1-w_0^2}{i\slash{p} -w_0^{N_s}(1-w_0^2) -(1-w_0^2) \, \Sigma_q (p)}
\nonumber \\ 
&=& \frac{1-w_0^2}{i\slash{p} \, Z_2^{-1} + m_{res}^{(1)}} \, Z_w \, , 
\label{eq:o-loop}
\end{eqnarray}
from which Eq.~(\ref{eq:mres}) follows. More details on these expressions 
concerning the one-loop self-energy can be found in \cite{Capitani:2006kw}.  
Here we only remind that we call $\bar g^2 = (g_0^2/16 \pi^2)\, C_F$ 
(with $C_F=4/3$ for QCD) and that $Z_2$ is the quark wave function 
renormalization factor, while $Z_w$ generates an additive renormalization to 
$w_0$ and hence to the domain-wall height $M$ \cite{Aoki:1998vv}. 
We can see that the mass correction term of Eq.~(\ref{eq:mres}) vanishes 
when the theory describes exact chiral fermions, but becomes nonzero when 
computations are done at any finite $N_s$. In this case $\Sigma_0$ generates 
a finite additive renormalization to the quark mass, which is a measure of 
chiral violations. We associate the perturbative critical mass $m_{res}$, 
which defines the chiral limit when no explicit mass term appears in the 
Lagrangian, with the residual mass which in Monte Carlo simulations is derived 
from the symmetry-breaking term in the axial Ward identities.

At one loop two diagrams enter in the calculation of the residual mass,
the half-circle (or sunset) and the tadpole diagrams, which at order zero
in $p$ contribute to $\Sigma_0$. The result of the tadpole diagram can be 
given in a simple form, and is equal to
\begin{widetext}
\begin{equation}
T_d^{(0)} = T_l^{(0)} \,\frac{1-w_0^2}{(1-w_0^{2N_s})^2} \,  
   \Big( N_s \, (1 + w_0^{2(N_s+1)}) \, w_0^{N_s-1} 
       - 2 \, w_0^{N_s+1} \, \frac{1-w_0^{2N_s}}{1-w_0^2} \Big) ,
\label{eq:tad-sigma0}
\end{equation}
\end{widetext}
where $T_l^{(0)}$ is (up to a sign) the result of the tadpole of order zero
for Wilson fermions: 
\begin{equation}
T_l^{(0)} = 40.517749 - 8 \pi^2 Z_0 \, (1-\lambda) 
\end{equation}
for the L\"uscher-Weisz action,
\begin{equation}
T_l^{(0)} = 29.927709 - 8 \pi^2 Z_0 \, (1-\lambda)  
\end{equation}
for the Iwasaki action and
\begin{equation}
T_l^{(0)} = 19.715871 - 8 \pi^2 Z_0 \, (1-\lambda)  
\end{equation}
for the DBW2 action, where $Z_0=0.154933390231\ldots$ is a well-known integral 
\cite{Capitani:2002mp}. It is useful to remind that for the plaquette action 
one had instead $T_l^{(0)} = 48.932201 - 8 \pi^2 Z_0 (1-\lambda)$. 
The domain-wall result assumes the simple form of Eq.~(\ref{eq:tad-sigma0}) 
because the tadpole diagram is diagonal in the indices of the extra dimension,
and since the loop integration involves only the gluon propagator, which can
be factorized out, for any given pair of $N_s$ and $M$ the tadpole results 
for all gauge actions are just proportional to $T_l^{(0)}$. We give here 
(in Table \ref{tab:sigma0taddb}) the explicit numbers only for the DBW2 action,
which is presently the most used in numerical simulations of domain-wall QCD; 
for the L\"uscher-Weisz and Iwasaki actions a simple multiplication then 
suffices.

The results for the half-circle diagram of $\Sigma_0$ are shown 
(in Feynman gauge) in Tables \ref{tab:sigma0hclw}, \ref{tab:sigma0hciw} 
and \ref{tab:sigma0hcdb} for the L\"uscher-Weisz, Iwasaki and DBW2 action
respectively. In addition to running the standard numerical integration in 6 
dimensions, we have also redone the computation of this diagram by hand, 
including the calculation of the gamma algebra and the explicit exact 
evaluation of the sums over the fifth-dimensional indices. The resulting 
expressions are then four-dimensional. This provides a rather strong check 
of our calculations, and also saves 2 dimensions in the numerical integration.

\begin{figure}
\includegraphics[width=9.1cm]{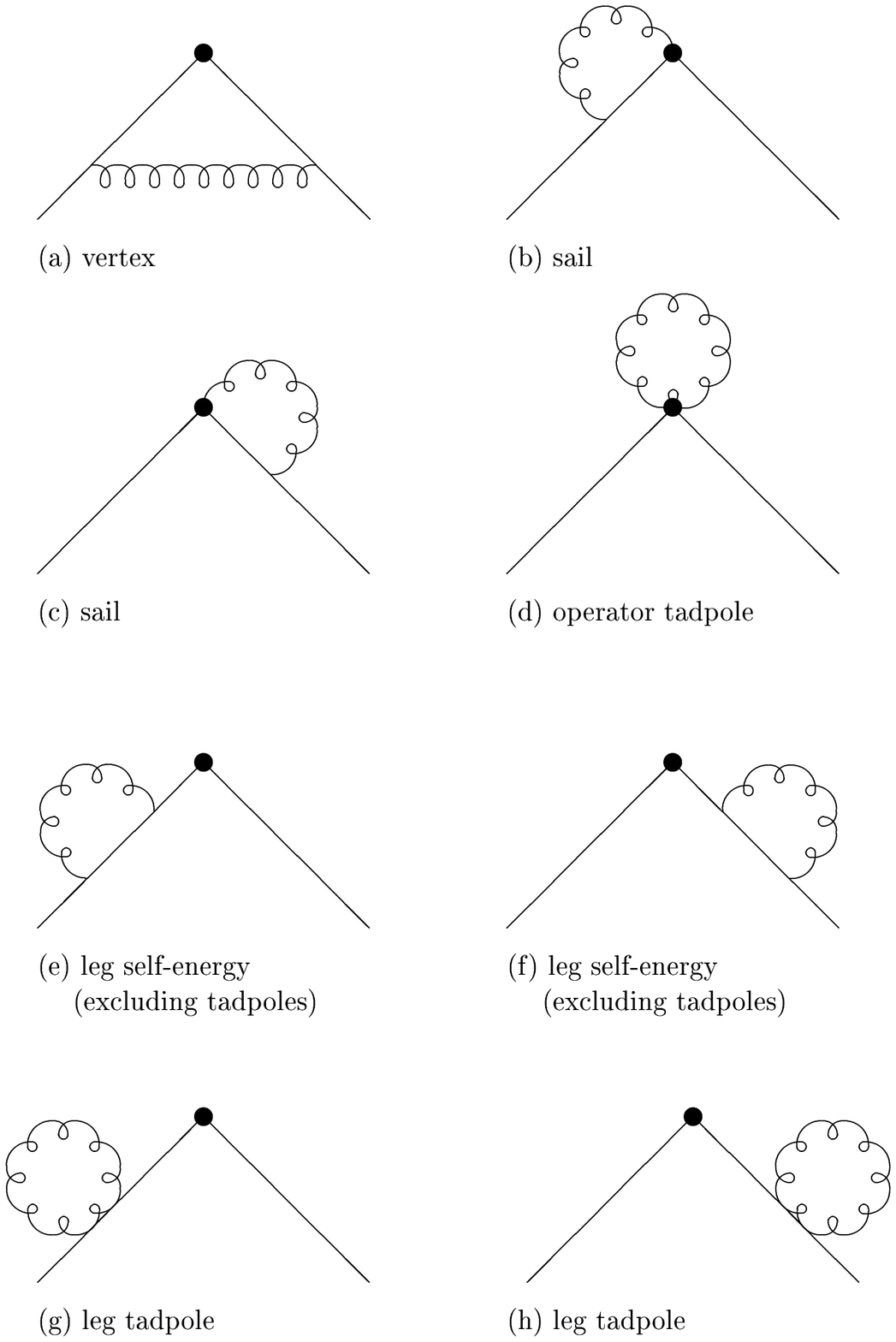}%
\caption{\label{fig:diagrams}The diagrams needed for the one-loop 
renormalization of the lattice operators.}
\end{figure}

\begin{table*}[htp]
\caption{\label{tab:sigma0taddb}
Coefficient of $\bar g^2$ for the tadpole contribution to $\Sigma_0$, 
Eq.~(\ref{eq:tad-sigma0}), in Feynman gauge for the DBW2 gauge action. 
For the Iwasaki action all entries have to be multiplied by $1.51795$, and for 
the L\"uscher-Weisz action all entries have to be multiplied by $2.05508$.}
\begin{ruledtabular}    
\begin{tabular}{|c|rrrrrrrrr|} 
$M$ & $N_s=8$ & $N_s=12$ & $N_s=16$ & $N_s=20$ & $N_s=24$ & $N_s=28$ 
    & $N_s=32$ & $N_s=48$ & $N_s=\infty$  
      \vspace{0.05cm} \\ \hline \vspace{-0.3cm} \\  
  0.1   &  6.08558  &  6.84244  &  6.79166  &  6.17209  &  5.26321
        &  4.28400  &  3.36856  &  1.04552  & 0 \\
  0.2   &  7.38915  &  5.23011  &  3.11540  &  1.68275  &  0.85663
        &  0.41950  &  0.19994  &  0.00879  & 0 \\
  0.3   &  5.08323  &  2.00488  &  0.67208  &  0.20721  &  0.06076
        &  0.01723  &  0.00477  &  0.00002  & 0 \\
  0.4   &  2.43021  &  0.49784  &  0.08825  &  0.01451  &  0.00228
        &  0.00035  &  0.00005  &  0.00000  & 0 \\
  0.5   &  0.84720  &  0.08183  &  0.00692  &  0.00055  &  0.00004
        &  0.00000  &  0.00000  &  0.00000  & 0 \\
  0.6   &  0.20674  &  0.00807  &  0.00028  &  0.00001  &  0.00000
        &  0.00000  &  0.00000  &  0.00000  & 0 \\
  0.7   &  0.03061  &  0.00038  &  0.00000  &  0.00000  &  0.00000
        &  0.00000  &  0.00000  &  0.00000  & 0 \\
  0.8   &  0.00192  &  0.00000  &  0.00000  &  0.00000  &  0.00000
        &  0.00000  &  0.00000  &  0.00000  & 0 \\
  0.9   &  0.00002  &  0.00000  &  0.00000  &  0.00000  &  0.00000
        &  0.00000  &  0.00000  &  0.00000  & 0 \\
  1.0   &  0.00000  &  0.00000  &  0.00000  &  0.00000  &  0.00000
        &  0.00000  &  0.00000  &  0.00000  & 0 \\
  1.1   & -0.00002  &  0.00000  &  0.00000  &  0.00000  &  0.00000
        &  0.00000  &  0.00000  &  0.00000  & 0 \\
  1.2   & -0.00192  &  0.00000  &  0.00000  &  0.00000  &  0.00000
        &  0.00000  &  0.00000  &  0.00000  & 0 \\
  1.3   & -0.03061  & -0.00038  &  0.00000  &  0.00000  &  0.00000
        &  0.00000  &  0.00000  &  0.00000  & 0 \\
  1.4   & -0.20674  & -0.00807  & -0.00028  & -0.00001  &  0.00000
        &  0.00000  &  0.00000  &  0.00000  & 0 \\
  1.5   & -0.84720  & -0.08183  & -0.00692  & -0.00055  & -0.00004
        &  0.00000  &  0.00000  &  0.00000  & 0 \\
  1.6   & -2.43021  & -0.49784  & -0.08825  & -0.01451  & -0.00228
        & -0.00035  & -0.00005  &  0.00000  & 0 \\
  1.7   & -5.08323  & -2.00488  & -0.67208  & -0.20721  & -0.06076
        & -0.01723  & -0.00477  & -0.00002  & 0 \\
  1.8   & -7.38915  & -5.23011  & -3.11540  & -1.68275  & -0.85663  
        & -0.41950  & -0.19994  & -0.00879  & 0 \\
  1.9   & -6.08558  & -6.84244  & -6.79166  & -6.17209  & -5.26321
        & -4.28400  & -3.36856  & -1.04552  & 0 \\ \hline    
\end{tabular}
\end{ruledtabular}
\end{table*}

\begin{table*}[htp]
\caption{\label{tab:sigma0hclw}
Coefficient of $\bar g^2$ for the contribution of the half-circle diagram 
to $\Sigma_0$, in Feynman gauge for the L\"uscher-Weisz gauge action.}
\begin{ruledtabular}    
\begin{tabular}{|c|rrrrrrrrr|} 
$M$ & $N_s=8$ & $N_s=12$ & $N_s=16$ & $N_s=20$ & $N_s=24$ & $N_s=28$ 
    & $N_s=32$ & $N_s=48$ & $N_s=\infty$  
      \vspace{0.05cm} \\ \hline \vspace{-0.3cm} \\  
  0.1   &  1.85305  &  1.69896  &  1.41789  &  1.12004  &  0.85404
        &  0.63588  &  0.46559  &  0.12125  & 0 \\
  0.2   &  2.33907  &  1.42243  &  0.76566  &  0.38732  &  0.18868
        &  0.08958  &  0.04173  &  0.00173  & 0 \\
  0.3   &  1.83151  &  0.67820  &  0.21995  &  0.06655  &  0.01928
        &  0.00542  &  0.00147  &  0.00001  & 0 \\
  0.4   &  1.08252  &  0.22556  &  0.04045  &  0.00670  &  0.00104
        &  0.00015  &  0.00002  &  0.00000  & 0 \\
  0.5   &  0.50879  &  0.05403  &  0.00482  &  0.00038  &  0.00003
        &  0.00000  &  0.00000  &  0.00000  & 0 \\
  0.6   &  0.19223  &  0.00935  &  0.00036  &  0.00001  &  0.00000
        &  0.00000  &  0.00000  &  0.00000  & 0 \\
  0.7   &  0.05966  &  0.00129  &  0.00002  &  0.00000  &  0.00000
        &  0.00000  &  0.00000  &  0.00000  & 0 \\
  0.8   &  0.01726  &  0.00025  &  0.00000  &  0.00000  &  0.00000
        &  0.00000  &  0.00000  &  0.00000  & 0 \\
  0.9   &  0.00635  &  0.00010  &  0.00000  &  0.00000  &  0.00000
        &  0.00000  &  0.00000  &  0.00000  & 0 \\
  1.0   &  0.00339  &  0.00007  &  0.00000  &  0.00000  &  0.00000
        &  0.00000  &  0.00000  &  0.00000  & 0 \\
  1.1   &  0.00229  &  0.00005  &  0.00000  &  0.00000  &  0.00000
        &  0.00000  &  0.00000  &  0.00000  & 0 \\
  1.2   &  0.00268  &  0.00005  &  0.00000  &  0.00000  &  0.00000
        &  0.00000  &  0.00000  &  0.00000  & 0 \\
  1.3   &  0.01527  &  0.00021  &  0.00000  &  0.00000  &  0.00000
        &  0.00000  &  0.00000  &  0.00000  & 0 \\
  1.4   &  0.08659  &  0.00335  &  0.00011  &  0.00000  &  0.00000
        &  0.00000  &  0.00000  &  0.00000  & 0 \\
  1.5   &  0.32093  &  0.03003  &  0.00249  &  0.00018  &  0.00001
        &  0.00000  &  0.00000  &  0.00000  & 0 \\
  1.6   &  0.84348  &  0.16170  &  0.02776  &  0.00448  &  0.00067
        &  0.00010  &  0.00001  &  0.00000  & 0 \\
  1.7   &  1.65276  &  0.57478  &  0.18073  &  0.05366  &  0.01535   
        &  0.00427  &  0.00115  &  0.00000  & 0 \\
  1.8   &  2.37106  &  1.36227  &  0.71569  &  0.35718  &  0.17245
        &  0.08134  &  0.03770  &  0.00154  & 0 \\
  1.9   &  2.08894  &  1.80335  &  1.46797  &  1.14614  &  0.86894
        &  0.64513  &  0.47169  &  0.12268  & 0 \\ \hline
\end{tabular}
\end{ruledtabular}
\end{table*}

\begin{table*}[htp]
\caption{\label{tab:sigma0hciw}
Coefficient of $\bar g^2$ for the contribution of the half-circle diagram 
to $\Sigma_0$, in Feynman gauge for the Iwasaki gauge action.}
\begin{ruledtabular}    
\begin{tabular}{|c|rrrrrrrrr|} 
$M$ & $N_s=8$ & $N_s=12$ & $N_s=16$ & $N_s=20$ & $N_s=24$ & $N_s=28$ 
    & $N_s=32$ & $N_s=48$ & $N_s=\infty$  
      \vspace{0.05cm} \\ \hline \vspace{-0.3cm} \\  
  0.1   &  0.50821  &  0.21779  & -0.03162  & -0.18502  & -0.25188
        & -0.26034  & -0.23689  & -0.09532  & 0 \\
  0.2   &  0.63587  &  0.22998  &  0.06006  &  0.00769  & -0.00410
        & -0.00467  & -0.00314  & -0.00024  & 0 \\
  0.3   &  0.57859  &  0.18254  &  0.05361  &  0.01524  &  0.00423
        &  0.00115  &  0.00029  &  0.00000  & 0 \\
  0.4   &  0.41985  &  0.08688  &  0.01562  &  0.00259  &  0.00039
        &  0.00005  &  0.00001  &  0.00000  & 0 \\
  0.5   &  0.23861  &  0.02652  &  0.00242  &  0.00019  &  0.00001
        &  0.00000  &  0.00000  &  0.00000  & 0 \\
  0.6   &  0.10593  &  0.00549  &  0.00022  &  0.00001  &  0.00000
        &  0.00000  &  0.00000  &  0.00000  & 0 \\
  0.7   &  0.03747  &  0.00086  &  0.00002  &  0.00000  &  0.00000
        &  0.00000  &  0.00000  &  0.00000  & 0 \\
  0.8   &  0.01169  &  0.00017  &  0.00000  &  0.00000  &  0.00000
        &  0.00000  &  0.00000  &  0.00000  & 0 \\
  0.9   &  0.00428  &  0.00007  &  0.00000  &  0.00000  &  0.00000
        &  0.00000  &  0.00000  &  0.00000  & 0 \\
  1.0   &  0.00222  &  0.00004  &  0.00000  &  0.00000  &  0.00000
        &  0.00000  &  0.00000  &  0.00000  & 0 \\
  1.1   &  0.00146  &  0.00003  &  0.00000  &  0.00000  &  0.00000
        &  0.00000  &  0.00000  &  0.00000  & 0 \\
  1.2   &  0.00218  &  0.00003  &  0.00000  &  0.00000  &  0.00000
        &  0.00000  &  0.00000  &  0.00000  & 0 \\
  1.3   &  0.01789  &  0.00023  &  0.00000  &  0.00000  &  0.00000
        &  0.00000  &  0.00000  &  0.00000  & 0 \\
  1.4   &  0.11072  &  0.00428  &  0.00014  &  0.00000  &  0.00000
        &  0.00000  &  0.00000  &  0.00000  & 0 \\
  1.5   &  0.43174  &  0.04077  &  0.00340  &  0.00025  &  0.00002
        &  0.00000  &  0.00000  &  0.00000  & 0 \\
  1.6   &  1.18690  &  0.23278  &  0.04042  &  0.00656  &  0.00100
        &  0.00015  &  0.00002  &  0.00000  & 0 \\
  1.7   &  2.41466  &  0.88093  &  0.28421  &  0.08571  &  0.02477
        &  0.00695  &  0.00189  &  0.00001  & 0 \\
  1.8   &  3.52723  &  2.20554  &  1.22505  &  0.63440  &  0.31424
        &  0.15100  &  0.07098  &  0.00301  & 0 \\
  1.9   &  3.06494  &  2.94845  &  2.63010  &  2.21616  &  1.78901
        &  1.39824  &  1.06623  &  0.30874  & 0 \\ \hline
\end{tabular}
\end{ruledtabular}
\end{table*}

\begin{table*}[htp]
\caption{\label{tab:sigma0hcdb}
Coefficient of $\bar g^2$ for the contribution of the half-circle diagram 
to $\Sigma_0$, in Feynman gauge for the DBW2 gauge action.}
\begin{ruledtabular}    
\begin{tabular}{|c|rrrrrrrrr|} 
$M$ & $N_s=8$ & $N_s=12$ & $N_s=16$ & $N_s=20$ & $N_s=24$ & $N_s=28$ 
    & $N_s=32$ & $N_s=48$ & $N_s=\infty$  
      \vspace{0.05cm} \\ \hline \vspace{-0.3cm} \\  
  0.1   & -1.10030  & -1.53568  & -1.73593  & -1.71261  & -1.54242
        & -1.30396  & -1.05364  & -0.34630  & 0 \\
  0.2   & -1.38867  & -1.17537  & -0.76769  & -0.43649  & -0.22928
        & -0.11464  & -0.05545  & -0.00253  & 0 \\
  0.3   & -0.90559  & -0.40170  & -0.14203  & -0.04505  & -0.01344
        & -0.00386  & -0.00110  & -0.00001  & 0 \\
  0.4   & -0.36485  & -0.07724  & -0.01379  & -0.00227  & -0.00038
        & -0.00006  & -0.00001  &  0.00000  & 0 \\
  0.5   & -0.08232  & -0.00635  & -0.00045  & -0.00004  &  0.00000
        &  0.00000  &  0.00000  &  0.00000  & 0 \\
  0.6   &  0.00264  &  0.00080  &  0.00004  &  0.00000  &  0.00000
        &  0.00000  &  0.00000  &  0.00000  & 0 \\
  0.7   &  0.01059  &  0.00032  &  0.00001  &  0.00000  &  0.00000
        &  0.00000  &  0.00000  &  0.00000  & 0 \\
  0.8   &  0.00494  &  0.00007  &  0.00000  &  0.00000  &  0.00000
        &  0.00000  &  0.00000  &  0.00000  & 0 \\
  0.9   &  0.00188  &  0.00003  &  0.00000  &  0.00000  &  0.00000
        &  0.00000  &  0.00000  &  0.00000  & 0 \\
  1.0   &  0.00092  &  0.00002  &  0.00000  &  0.00000  &  0.00000      
        &  0.00000  &  0.00000  &  0.00000  & 0 \\
  1.1   &  0.00059  &  0.00001  &  0.00000  &  0.00000  &  0.00000
        &  0.00000  &  0.00000  &  0.00000  & 0 \\
  1.2   &  0.00169  &  0.00001  &  0.00000  &  0.00000  &  0.00000
        &  0.00000  &  0.00000  &  0.00000  & 0 \\
  1.3   &  0.02083  &  0.00025  &  0.00000  &  0.00000  &  0.00000
        &  0.00000  &  0.00000  &  0.00000  & 0 \\
  1.4   &  0.13739  &  0.00532  &  0.00017  &  0.00001  &  0.00000
        &  0.00000  &  0.00000  &  0.00000  & 0 \\
  1.5   &  0.55390  &  0.05271  &  0.00442  &  0.00033  &  0.00002
        &  0.00000  &  0.00000  &  0.00000  & 0 \\
  1.6   &  1.56471  &  0.31178  &  0.05455  &  0.00890  &  0.00137
        &  0.00020  &  0.00003  &  0.00000  & 0 \\
  1.7   &  3.24961  &  1.22111  &  0.39990  &  0.12166  &  0.03537
        &  0.00997  &  0.00273  &  0.00001  & 0 \\
  1.8   &  4.78426  &  3.14023  &  1.79449  &  0.94576  &  0.47393
        &  0.22960  &  0.10857  &  0.00468  & 0 \\
  1.9   &  4.11043  &  4.20697  &  3.92344  &  3.41545  &  2.82481
        &  2.24857  &  1.73889  &  0.52014  & 0 \\ \hline
\end{tabular}
\end{ruledtabular}
\end{table*}

\begin{table*}[htp]
\caption{\label{tab:sigma0lw}
Coefficient of $\bar g^2$ for the complete result of $\Sigma_0$, in Feynman 
gauge for the L\"uscher-Weisz gauge action.}
\begin{ruledtabular}    
\begin{tabular}{|c|rrrrrrrrr|} 
$M$ & $N_s=8$ & $N_s=12$ & $N_s=16$ & $N_s=20$ & $N_s=24$ & $N_s=28$ 
    & $N_s=32$ & $N_s=48$ & $N_s=\infty$  
      \vspace{0.05cm} \\ \hline \vspace{-0.3cm} \\  
  0.1   & 14.35942  & 15.76075  & 15.37532  & 13.80421  & 11.67038
        &  9.43986  &  7.38826  &  5.42456  & 0 \\
  0.2   & 17.52439  & 12.17074  &  7.16806  &  3.84552  &  1.94912
        &  0.95169  &  0.45262  &  0.19370  & 0 \\
  0.3   & 12.27798  &  4.79840  &  1.60113  &  0.49238  &  0.14415
        &  0.04083  &  0.01128  &  0.00267  & 0 \\
  0.4   &  6.07680  &  1.24867  &  0.22182  &  0.03653  &  0.00572
        &  0.00087  &  0.00013  &  0.00002  & 0 \\
  0.5   &  2.24985  &  0.22220  &  0.01904  &  0.00150  &  0.00011
        &  0.00001  &  0.00000  &  0.00000  & 0 \\
  0.6   &  0.61709  &  0.02594  &  0.00093  &  0.00003  &  0.00000
        &  0.00000  &  0.00000  &  0.00000  & 0 \\
  0.7   &  0.12258  &  0.00206  &  0.00003  &  0.00000  &  0.00000
        &  0.00000  &  0.00000  &  0.00000  & 0 \\
  0.8   &  0.02120  &  0.00025  &  0.00000  &  0.00000  &  0.00000
        &  0.00000  &  0.00000  &  0.00000  & 0 \\
  0.9   &  0.00638  &  0.00010  &  0.00000  &  0.00000  &  0.00000
        &  0.00000  &  0.00000  &  0.00000  & 0 \\
  1.0   &  0.00339  &  0.00007  &  0.00000  &  0.00000  &  0.00000
        &  0.00000  &  0.00000  &  0.00000  & 0 \\
  1.1   &  0.00226  &  0.00005  &  0.00000  &  0.00000  &  0.00000
        &  0.00000  &  0.00000  &  0.00000  & 0 \\
  1.2   & -0.00126  &  0.00005  &  0.00000  &  0.00000  &  0.00000
        &  0.00000  &  0.00000  &  0.00000  & 0 \\
  1.3   & -0.04765  & -0.00056  &  0.00000  &  0.00000  &  0.00000
        &  0.00000  &  0.00000  &  0.00000  & 0 \\
  1.4   & -0.33827  & -0.01324  & -0.00046  & -0.00001  &  0.00000
        &  0.00000  &  0.00000  &  0.00000  & 0 \\
  1.5   & -1.42013  & -0.13814  & -0.01173  & -0.00094  & -0.00007
        & -0.00001  &  0.00000  &  0.00000  & 0 \\
  1.6   & -4.15080  & -0.86142  & -0.15360  & -0.02535  & -0.00401
        & -0.00062  & -0.00009  & -0.00002  & 0 \\
  1.7   & -8.79371  & -3.54541  & -1.20046  & -0.37218  & -0.10952
        & -0.03114  & -0.00866  & -0.00266  & 0 \\
  1.8   &-12.81426  & -9.38605  & -5.68671  & -3.10102  & -1.58799
        & -0.78077  & -0.37320  & -0.19043  & 0 \\
  1.9   &-10.41742  &-12.25844  &-12.48946  &-11.53802  & -9.94739
        & -8.15885  & -6.45097  & -5.18063  & 0 \\ \hline    
\end{tabular}
\end{ruledtabular}
\end{table*}

\begin{table*}[htp]
\caption{\label{tab:sigma0iw}
Coefficient of $\bar g^2$ for the complete result of $\Sigma_0$, in Feynman 
gauge for the Iwasaki gauge action.}
\begin{ruledtabular}    
\begin{tabular}{|c|rrrrrrrrr|} 
$M$ & $N_s=8$ & $N_s=12$ & $N_s=16$ & $N_s=20$ & $N_s=24$ & $N_s=28$ 
    & $N_s=32$ & $N_s=48$ & $N_s=\infty$  
      \vspace{0.05cm} \\ \hline \vspace{-0.3cm} \\  
  0.1   &  9.74581  & 10.60427  & 10.27779  &  9.18391  &  7.73742
        &  6.24256  &  4.87641  &  3.82187  & 0 \\
  0.2   & 11.85223  &  8.16903  &  4.78908  &  2.56202  &  1.29622
        &  0.63211  &  0.30036  &  0.14156  & 0 \\
  0.3   &  8.29469  &  3.22585  &  1.07380  &  0.32977  &  0.09646
        &  0.02731  &  0.00753  &  0.00197  & 0 \\
  0.4   &  4.10879  &  0.84259  &  0.14958  &  0.02462  &  0.00385
        &  0.00058  &  0.00009  &  0.00001  & 0 \\
  0.5   &  1.52461  &  0.15073  &  0.01293  &  0.00102  &  0.00008
        &  0.00001  &  0.00000  &  0.00000  & 0 \\
  0.6   &  0.41974  &  0.01774  &  0.00064  &  0.00002  &  0.00000
        &  0.00000  &  0.00000  &  0.00000  & 0 \\
  0.7   &  0.08394  &  0.00143  &  0.00002  &  0.00000  &  0.00000
        &  0.00000  &  0.00000  &  0.00000  & 0 \\
  0.8   &  0.01460  &  0.00017  &  0.00000  &  0.00000  &  0.00000
        &  0.00000  &  0.00000  &  0.00000  & 0 \\
  0.9   &  0.00431  &  0.00007  &  0.00000  &  0.00000  &  0.00000
        &  0.00000  &  0.00000  &  0.00000  & 0 \\
  1.0   &  0.00222  &  0.00004  &  0.00000  &  0.00000  &  0.00000
        &  0.00000  &  0.00000  &  0.00000  & 0 \\
  1.1   &  0.00144  &  0.00003  &  0.00000  &  0.00000  &  0.00000
        &  0.00000  &  0.00000  &  0.00000  & 0 \\
  1.2   & -0.00073  &  0.00003  &  0.00000  &  0.00000  &  0.00000
        &  0.00000  &  0.00000  &  0.00000  & 0 \\
  1.3   & -0.02859  & -0.00034  &  0.00000  &  0.00000  &  0.00000
        &  0.00000  &  0.00000  &  0.00000  & 0 \\
  1.4   & -0.20310  & -0.00797  & -0.00028  & -0.00001  &  0.00000
        &  0.00000  &  0.00000  &  0.00000  & 0 \\
  1.5   & -0.85426  & -0.08344  & -0.00710  & -0.00057  & -0.00004
        &  0.00000  &  0.00000  &  0.00000  & 0 \\
  1.6   & -2.50204  & -0.52292  & -0.09354  & -0.01547  & -0.00246
        & -0.00038  & -0.00006  & -0.00001  & 0 \\
  1.7   & -5.30143  & -2.16238  & -0.73598  & -0.22883  & -0.06746
        & -0.01920  & -0.00535  & -0.00196  & 0 \\
  1.8   & -7.68914  & -5.73352  & -3.50397  & -1.91993  & -0.98607
        & -0.48578  & -0.23252  & -0.13878  & 0 \\
  1.9   & -6.17267  & -7.43803  & -7.67930  & -7.15278  & -6.20028
        & -5.10466  & -4.04708  & -3.60846  & 0 \\ \hline    
\end{tabular}
\end{ruledtabular}
\end{table*}

\begin{table*}[htp]
\caption{\label{tab:sigma0db}
Coefficient of $\bar g^2$ for the complete result of $\Sigma_0$, in Feynman 
gauge for the DBW2 gauge action.}
\begin{ruledtabular}    
\begin{tabular}{|c|rrrrrrrrr|} 
$M$ & $N_s=8$ & $N_s=12$ & $N_s=16$ & $N_s=20$ & $N_s=24$ & $N_s=28$ 
    & $N_s=32$ & $N_s=48$ & $N_s=\infty$  
      \vspace{0.05cm} \\ \hline \vspace{-0.3cm} \\  
  0.1   &  4.98528  &  5.30676  &  5.05573  &  4.45949  &  3.72080
        &  2.98004  &  2.31491  &  2.23429  & 0 \\
  0.2   &  6.00048  &  4.05474  &  2.34771  &  1.24626  &  0.62735
        &  0.30486  &  0.14449  &  0.09088  & 0 \\
  0.3   &  4.17765  &  1.60318  &  0.53006  &  0.16216  &  0.04732
        &  0.01337  &  0.00368  &  0.00129  & 0 \\
  0.4   &  2.06536  &  0.42060  &  0.07446  &  0.01224  &  0.00190
        &  0.00028  &  0.00004  &  0.00001  & 0 \\
  0.5   &  0.76488  &  0.07548  &  0.00647  &  0.00050  &  0.00004
        &  0.00000  &  0.00000  &  0.00000  & 0 \\
  0.6   &  0.20938  &  0.00887  &  0.00032  &  0.00001  &  0.00000
        &  0.00000  &  0.00000  &  0.00000  & 0 \\
  0.7   &  0.04121  &  0.00070  &  0.00001  &  0.00000  &  0.00000
        &  0.00000  &  0.00000  &  0.00000  & 0 \\
  0.8   &  0.00686  &  0.00008  &  0.00000  &  0.00000  &  0.00000
        &  0.00000  &  0.00000  &  0.00000  & 0 \\
  0.9   &  0.00189  &  0.00003  &  0.00000  &  0.00000  &  0.00000
        &  0.00000  &  0.00000  &  0.00000  & 0 \\
  1.0   &  0.00092  &  0.00002  &  0.00000  &  0.00000  &  0.00000
        &  0.00000  &  0.00000  &  0.00000  & 0 \\
  1.1   &  0.00057  &  0.00001  &  0.00000  &  0.00000  &  0.00000
        &  0.00000  &  0.00000  &  0.00000  & 0 \\
  1.2   & -0.00023  &  0.00001  &  0.00000  &  0.00000  &  0.00000
        &  0.00000  &  0.00000  &  0.00000  & 0 \\
  1.3   & -0.00978  & -0.00012  &  0.00000  &  0.00000  &  0.00000
        &  0.00000  &  0.00000  &  0.00000  & 0 \\
  1.4   & -0.06935  & -0.00275  & -0.00010  &  0.00000  &  0.00000
        &  0.00000  &  0.00000  &  0.00000  & 0 \\
  1.5   & -0.29330  & -0.02912  & -0.00250  & -0.00021  & -0.00002
        &  0.00000  &  0.00000  &  0.00000  & 0 \\
  1.6   & -0.86550  & -0.18607  & -0.03370  & -0.00561  & -0.00091
        & -0.00014  & -0.00002  & -0.00001  & 0 \\
  1.7   & -1.83363  & -0.78377  & -0.27219  & -0.08555  & -0.02539
        & -0.00726  & -0.00204  & -0.00129  & 0 \\
  1.8   & -2.60489  & -2.08988  & -1.32091  & -0.73700  & -0.38269
        & -0.18990  & -0.09137  & -0.08873  & 0 \\
  1.9   & -1.97515  & -2.63547  & -2.86822  & -2.75664  & -2.43840
        & -2.03543  & -1.62967  & -2.06044  & 0 \\ \hline    
\end{tabular}
\end{ruledtabular}
\end{table*}

\begin{table*}[htp]
\caption{\label{tab:residuall52lw}
Residual mass in lattice units at $\beta=5.2$, in Landau gauge 
for the L\"uscher-Weisz gauge action.} 
\begin{ruledtabular}    
\begin{tabular}{|c|rrrrrrrrr|} 
$M$ & $N_s=8$ & $N_s=12$ & $N_s=16$ & $N_s=20$ & $N_s=24$ & $N_s=28$ 
    & $N_s=32$ & $N_s=48$ & $N_s=\infty$  
      \vspace{0.05cm} \\ \hline \vspace{-0.3cm} \\  
  0.1   & -0.21605  & -0.20260  & -0.18145  & -0.15496  & -0.12696
        & -0.10057  & -0.07757  & -0.04763  & 0 \\
  0.2   & -0.22551  & -0.14044  & -0.07859  & -0.04098  & -0.02040
        & -0.00984  & -0.00464  & -0.00149  & 0 \\
  0.3   & -0.14568  & -0.05280  & -0.01700  & -0.00512  & -0.00148
        & -0.00042  & -0.00011  & -0.00002  & 0 \\
  0.4   & -0.06850  & -0.01331  & -0.00230  & -0.00037  & -0.00006
        & -0.00001  &  0.00000  &  0.00000  & 0 \\
  0.5   & -0.02438  & -0.00231  & -0.00019  & -0.00002  &  0.00000
        &  0.00000  &  0.00000  &  0.00000  & 0 \\
  0.6   & -0.00646  & -0.00026  & -0.00001  &  0.00000  &  0.00000
        &  0.00000  &  0.00000  &  0.00000  & 0 \\
  0.7   & -0.00124  & -0.00002  &  0.00000  &  0.00000  &  0.00000
        &  0.00000  &  0.00000  &  0.00000  & 0 \\
  0.8   & -0.00021  &  0.00000  &  0.00000  &  0.00000  &  0.00000
        &  0.00000  &  0.00000  &  0.00000  & 0 \\
  0.9   & -0.00006  &  0.00000  &  0.00000  &  0.00000  &  0.00000
        &  0.00000  &  0.00000  &  0.00000  & 0 \\
  1.0   & -0.00003  &  0.00000  &  0.00000  &  0.00000  &  0.00000
        &  0.00000  &  0.00000  &  0.00000  & 0 \\
  1.1   & -0.00002  &  0.00000  &  0.00000  &  0.00000  &  0.00000
        &  0.00000  &  0.00000  &  0.00000  & 0 \\
  1.2   &  0.00001  &  0.00000  &  0.00000  &  0.00000  &  0.00000
        &  0.00000  &  0.00000  &  0.00000  & 0 \\
  1.3   &  0.00042  &  0.00001  &  0.00000  &  0.00000  &  0.00000
        &  0.00000  &  0.00000  &  0.00000  & 0 \\
  1.4   &  0.00285  &  0.00012  &  0.00000  &  0.00000  &  0.00000
        &  0.00000  &  0.00000  &  0.00000  & 0 \\
  1.5   &  0.01138  &  0.00120  &  0.00011  &  0.00001  &  0.00000
        &  0.00000  &  0.00000  &  0.00000  & 0 \\
  1.6   &  0.03114  &  0.00724  &  0.00135  &  0.00023  &  0.00004
        &  0.00001  &  0.00000  &  0.00000  & 0 \\
  1.7   &  0.05963  &  0.02849  &  0.01029  &  0.00330  &  0.00099
        &  0.00029  &  0.00008  &  0.00002  & 0 \\
  1.8   &  0.07020  &  0.06959  &  0.04664  &  0.02670  &  0.01406
        &  0.00704  &  0.00341  &  0.00123  & 0 \\
  1.9   &  0.02569  &  0.07052  &  0.09008  &  0.09195  &  0.08365
        &  0.07088  &  0.05726  &  0.03713  & 0 \\ \hline    
\end{tabular}
\end{ruledtabular}
\end{table*}

\begin{table*}[htp]
\caption{\label{tab:residuall52iw}
Residual mass in lattice units at $\beta=5.2$, in Landau gauge 
for the Iwasaki gauge action.} 
\begin{ruledtabular}    
\begin{tabular}{|c|rrrrrrrrr|} 
$M$ & $N_s=8$ & $N_s=12$ & $N_s=16$ & $N_s=20$ & $N_s=24$ & $N_s=28$ 
    & $N_s=32$ & $N_s=48$ & $N_s=\infty$  
      \vspace{0.05cm} \\ \hline \vspace{-0.3cm} \\  
  0.1   & -0.17110  & -0.15236  & -0.13179  & -0.10995  & -0.08865
        & -0.06942  & -0.05309  & -0.03202  & 0 \\
  0.2   & -0.17025  & -0.10145  & -0.05542  & -0.02847  & -0.01404
        & -0.00672  & -0.00315  & -0.00098  & 0 \\
  0.3   & -0.10687  & -0.03748  & -0.01187  & -0.00354  & -0.00102
        & -0.00028  & -0.00008  & -0.00001  & 0 \\
  0.4   & -0.04933  & -0.00936  & -0.00160  & -0.00026  & -0.00004
        & -0.00001  &  0.00000  &  0.00000  & 0 \\
  0.5   & -0.01731  & -0.00161  & -0.00013  & -0.00001  &  0.00000
        &  0.00000  &  0.00000  &  0.00000  & 0 \\
  0.6   & -0.00453  & -0.00018  & -0.00001  &  0.00000  &  0.00000
        &  0.00000  &  0.00000  &  0.00000  & 0 \\
  0.7   & -0.00086  & -0.00001  &  0.00000  &  0.00000  &  0.00000
        &  0.00000  &  0.00000  &  0.00000  & 0 \\
  0.8   & -0.00014  &  0.00000  &  0.00000  &  0.00000  &  0.00000
        &  0.00000  &  0.00000  &  0.00000  & 0 \\
  0.9   & -0.00004  &  0.00000  &  0.00000  &  0.00000  &  0.00000
        &  0.00000  &  0.00000  &  0.00000  & 0 \\
  1.0   & -0.00002  &  0.00000  &  0.00000  &  0.00000  &  0.00000
        &  0.00000  &  0.00000  &  0.00000  & 0 \\
  1.1   & -0.00001  &  0.00000  &  0.00000  &  0.00000  &  0.00000
        &  0.00000  &  0.00000  &  0.00000  & 0 \\
  1.2   &  0.00001  &  0.00000  &  0.00000  &  0.00000  &  0.00000
        &  0.00000  &  0.00000  &  0.00000  & 0 \\
  1.3   &  0.00023  &  0.00000  &  0.00000  &  0.00000  &  0.00000
        &  0.00000  &  0.00000  &  0.00000  & 0 \\
  1.4   &  0.00153  &  0.00007  &  0.00000  &  0.00000  &  0.00000
        &  0.00000  &  0.00000  &  0.00000  & 0 \\
  1.5   &  0.00586  &  0.00067  &  0.00006  &  0.00001  &  0.00000
        &  0.00000  &  0.00000  &  0.00000  & 0 \\
  1.6   &  0.01508  &  0.00394  &  0.00077  &  0.00013  &  0.00002
        &  0.00000  &  0.00000  &  0.00000  & 0 \\
  1.7   &  0.02561  &  0.01502  &  0.00577  &  0.00190  &  0.00058
        &  0.00017  &  0.00005  &  0.00001  & 0 \\
  1.8   &  0.02027  &  0.03401  &  0.02538  &  0.01519  &  0.00820
        &  0.00417  &  0.00204  &  0.00073  & 0 \\
  1.9   & -0.01566  &  0.02356  &  0.04321  &  0.04923  &  0.04715
        &  0.04113  &  0.03384  &  0.02181  & 0 \\ \hline    
\end{tabular}
\end{ruledtabular}
\end{table*}

\begin{table*}[htp]
\caption{\label{tab:residualf60db}
Residual mass in lattice units at $\beta=6$, in Feynman gauge
for the DBW2 gauge action.} 
\begin{ruledtabular}    
\begin{tabular}{|c|rrrrrrrrr|} 
$M$ & $N_s=8$ & $N_s=12$ & $N_s=16$ & $N_s=20$ & $N_s=24$ & $N_s=28$ 
    & $N_s=32$ & $N_s=48$ & $N_s=\infty$  
      \vspace{0.05cm} \\ \hline \vspace{-0.3cm} \\  
  0.1   & -0.12388  & -0.09847  & -0.07790  & -0.06075  & -0.04657
        & -0.03511  & -0.02607  & -0.02315  & 0 \\
  0.2   & -0.11106  & -0.05897  & -0.02996  & -0.01467  & -0.00700
        & -0.00327  & -0.00151  & -0.00088  & 0 \\
  0.3   & -0.06467  & -0.02060  & -0.00617  & -0.00178  & -0.00050
        & -0.00014  & -0.00004  & -0.00001  & 0 \\
  0.4   & -0.02819  & -0.00494  & -0.00081  & -0.00013  & -0.00002
        &  0.00000  &  0.00000  &  0.00000  & 0 \\
  0.5   & -0.00939  & -0.00082  & -0.00007  &  0.00000  &  0.00000
        &  0.00000  &  0.00000  &  0.00000  & 0 \\
  0.6   & -0.00232  & -0.00009  &  0.00000  &  0.00000  &  0.00000
        &  0.00000  &  0.00000  &  0.00000  & 0 \\
  0.7   & -0.00041  & -0.00001  &  0.00000  &  0.00000  &  0.00000
        &  0.00000  &  0.00000  &  0.00000  & 0 \\
  0.8   & -0.00006  &  0.00000  &  0.00000  &  0.00000  &  0.00000
        &  0.00000  &  0.00000  &  0.00000  & 0 \\
  0.9   & -0.00002  &  0.00000  &  0.00000  &  0.00000  &  0.00000
        &  0.00000  &  0.00000  &  0.00000  & 0 \\
  1.0   & -0.00001  &  0.00000  &  0.00000  &  0.00000  &  0.00000
        &  0.00000  &  0.00000  &  0.00000  & 0 \\
  1.1   &  0.00000  &  0.00000  &  0.00000  &  0.00000  &  0.00000
        &  0.00000  &  0.00000  &  0.00000  & 0 \\
  1.2   &  0.00000  &  0.00000  &  0.00000  &  0.00000  &  0.00000
        &  0.00000  &  0.00000  &  0.00000  & 0 \\
  1.3   &  0.00002  &  0.00000  &  0.00000  &  0.00000  &  0.00000
        &  0.00000  &  0.00000  &  0.00000  & 0 \\
  1.4   &  0.00004  &  0.00001  &  0.00000  &  0.00000  &  0.00000
        &  0.00000  &  0.00000  &  0.00000  & 0 \\
  1.5   & -0.00045  &  0.00006  &  0.00001  &  0.00000  &  0.00000 
        &  0.00000  &  0.00000  &  0.00000  & 0 \\
  1.6   & -0.00344  &  0.00018  &  0.00010  &  0.00002  &  0.00000
        &  0.00000  &  0.00000  &  0.00000  & 0 \\
  1.7   & -0.01392  & -0.00044  &  0.00060  &  0.00032  &  0.00012
        &  0.00004  &  0.00001  &  0.00001  & 0 \\
  1.8   & -0.03840  & -0.00709  &  0.00102  &  0.00207  &  0.00153
        &  0.00091  &  0.00049  &  0.00063  & 0 \\
  1.9   & -0.06511  & -0.03141  & -0.01099  &  0.00018  &  0.00543
        &  0.00724  &  0.00724  &  0.01312  & 0 \\ \hline    
\end{tabular}
\end{ruledtabular}
\end{table*}

\begin{table*}[htp]
\caption{\label{tab:residuall60db}
Residual mass in lattice units at $\beta=6$, in Landau gauge
for the DBW2 gauge action.} 
\begin{ruledtabular}    
\begin{tabular}{|c|rrrrrrrrr|} 
$M$ & $N_s=8$ & $N_s=12$ & $N_s=16$ & $N_s=20$ & $N_s=24$ & $N_s=28$ 
    & $N_s=32$ & $N_s=48$ & $N_s=\infty$  
      \vspace{0.05cm} \\ \hline \vspace{-0.3cm} \\  
  0.1   & -0.11900  & -0.09447  & -0.07482  & -0.05848  & -0.04493
        & -0.03395  & -0.02526  & -0.01492  & 0 \\
  0.2   & -0.10619  & -0.05648  & -0.02877  & -0.01412  & -0.00675
        & -0.00316  & -0.00145  & -0.00044  & 0 \\
  0.3   & -0.06178  & -0.01972  & -0.00592  & -0.00171  & -0.00048
        & -0.00013  & -0.00004  & -0.00001  & 0 \\
  0.4   & -0.02693  & -0.00473  & -0.00078  & -0.00012  & -0.00002
        &  0.00000  &  0.00000  &  0.00000  & 0 \\
  0.5   & -0.00898  & -0.00079  & -0.00006  &  0.00000  &  0.00000
        &  0.00000  &  0.00000  &  0.00000  & 0 \\
  0.6   & -0.00223  & -0.00009  &  0.00000  &  0.00000  &  0.00000
        &  0.00000  &  0.00000  &  0.00000  & 0 \\
  0.7   & -0.00040  & -0.00001  &  0.00000  &  0.00000  &  0.00000
        &  0.00000  &  0.00000  &  0.00000  & 0 \\
  0.8   & -0.00006  &  0.00000  &  0.00000  &  0.00000  &  0.00000
        &  0.00000  &  0.00000  &  0.00000  & 0 \\
  0.9   & -0.00002  &  0.00000  &  0.00000  &  0.00000  &  0.00000
        &  0.00000  &  0.00000  &  0.00000  & 0 \\
  1.0   & -0.00001  &  0.00000  &  0.00000  &  0.00000  &  0.00000
        &  0.00000  &  0.00000  &  0.00000  & 0 \\
  1.1   &  0.00000  &  0.00000  &  0.00000  &  0.00000  &  0.00000
        &  0.00000  &  0.00000  &  0.00000  & 0 \\
  1.2   &  0.00000  &  0.00000  &  0.00000  &  0.00000  &  0.00000
        &  0.00000  &  0.00000  &  0.00000  & 0 \\
  1.3   &  0.00004  &  0.00000  &  0.00000  &  0.00000  &  0.00000
        &  0.00000  &  0.00000  &  0.00000  & 0 \\
  1.4   &  0.00013  &  0.00001  &  0.00000  &  0.00000  &  0.00000  
        &  0.00000  &  0.00000  &  0.00000  & 0 \\
  1.5   & -0.00005  &  0.00010  &  0.00001  &  0.00000  &  0.00000
        &  0.00000  &  0.00000  &  0.00000  & 0 \\
  1.6   & -0.00219  &  0.00039  &  0.00014  &  0.00003  &  0.00001
        &  0.00000  &  0.00000  &  0.00000  & 0 \\
  1.7   & -0.01101  &  0.00043  &  0.00085  &  0.00039  &  0.00014
        &  0.00004  &  0.00001  &  0.00000  & 0 \\
  1.8   & -0.03342  & -0.00459  &  0.00221  &  0.00262  &  0.00178
        &  0.00102  &  0.00054  &  0.00019  & 0 \\
  1.9   & -0.05992  & -0.02729  & -0.00787  &  0.00247  &  0.00708
        &  0.00840  &  0.00805  &  0.00526  & 0 \\ \hline    
\end{tabular}
\end{ruledtabular}
\end{table*}

\begin{table*}[htp]
\caption{\label{tab:residualf52db}
Residual mass in lattice units at $\beta=5.2$, in Feynman gauge
for the DBW2 gauge action.} 
\begin{ruledtabular}    
\begin{tabular}{|c|rrrrrrrrr|} 
$M$ & $N_s=8$ & $N_s=12$ & $N_s=16$ & $N_s=20$ & $N_s=24$ & $N_s=28$ 
    & $N_s=32$ & $N_s=48$ & $N_s=\infty$  
      \vspace{0.05cm} \\ \hline \vspace{-0.3cm} \\  
  0.1   & -0.13036  & -0.10536  & -0.08446  & -0.06655  & -0.05141
        & -0.03898  & -0.02908  & -0.02605  & 0 \\
  0.2   & -0.11886  & -0.06424  & -0.03301  & -0.01629  & -0.00781
        & -0.00367  & -0.00169  & -0.00100  & 0 \\
  0.3   & -0.07010  & -0.02268  & -0.00686  & -0.00199  & -0.00056
        & -0.00015  & -0.00004  & -0.00001  & 0 \\
  0.4   & -0.03087  & -0.00549  & -0.00091  & -0.00014  & -0.00002
        &  0.00000  &  0.00000  &  0.00000  & 0 \\
  0.5   & -0.01038  & -0.00092  & -0.00007  & -0.00001  &  0.00000
        &  0.00000  &  0.00000  &  0.00000  & 0 \\
  0.6   & -0.00259  & -0.00010  &  0.00000  &  0.00000  &  0.00000
        &  0.00000  &  0.00000  &  0.00000  & 0 \\
  0.7   & -0.00046  & -0.00001  &  0.00000  &  0.00000  &  0.00000
        &  0.00000  &  0.00000  &  0.00000  & 0 \\
  0.8   & -0.00007  &  0.00000  &  0.00000  &  0.00000  &  0.00000
        &  0.00000  &  0.00000  &  0.00000  & 0 \\
  0.9   & -0.00002  &  0.00000  &  0.00000  &  0.00000  &  0.00000
        &  0.00000  &  0.00000  &  0.00000  & 0 \\
  1.0   & -0.00001  &  0.00000  &  0.00000  &  0.00000  &  0.00000
        &  0.00000  &  0.00000  &  0.00000  & 0 \\
  1.1   & -0.00001  &  0.00000  &  0.00000  &  0.00000  &  0.00000
        &  0.00000  &  0.00000  &  0.00000  & 0 \\
  1.2   &  0.00000  &  0.00000  &  0.00000  &  0.00000  &  0.00000
        &  0.00000  &  0.00000  &  0.00000  & 0 \\
  1.3   &  0.00004  &  0.00000  &  0.00000  &  0.00000  &  0.00000
        &  0.00000  &  0.00000  &  0.00000  & 0 \\
  1.4   &  0.00013  &  0.00001  &  0.00000  &  0.00000  &  0.00000
        &  0.00000  &  0.00000  &  0.00000  & 0 \\
  1.5   & -0.00007  &  0.00010  &  0.00001  &  0.00000  &  0.00000
        &  0.00000  &  0.00000  &  0.00000  & 0 \\
  1.6   & -0.00232  &  0.00042  &  0.00015  &  0.00003  &  0.00001
        &  0.00000  &  0.00000  &  0.00000  & 0 \\
  1.7   & -0.01154  &  0.00058  &  0.00096  &  0.00043  &  0.00015
        &  0.00005  &  0.00001  &  0.00001  & 0 \\
  1.8   & -0.03502  & -0.00438  &  0.00274  &  0.00303  &  0.00203
        &  0.00115  &  0.00060  &  0.00075  & 0 \\
  1.9   & -0.06255  & -0.02799  & -0.00726  &  0.00376  &  0.00860
        &  0.00989  &  0.00935  &  0.01579  & 0 \\ \hline    
\end{tabular}
\end{ruledtabular}
\end{table*}

\begin{table*}[htp]
\caption{\label{tab:residuall52db}
Residual mass in lattice units at $\beta=5.2$, in Landau gauge 
for the DBW2 gauge action.} 
\begin{ruledtabular}    
\begin{tabular}{|c|rrrrrrrrr|} 
$M$ & $N_s=8$ & $N_s=12$ & $N_s=16$ & $N_s=20$ & $N_s=24$ & $N_s=28$ 
    & $N_s=32$ & $N_s=48$ & $N_s=\infty$  
      \vspace{0.05cm} \\ \hline \vspace{-0.3cm} \\  
  0.1   & -0.12472  & -0.10075  & -0.08091  & -0.06392  & -0.04952
        & -0.03764  & -0.02814  & -0.01655  & 0 \\
  0.2   & -0.11324  & -0.06137  & -0.03163  & -0.01566  & -0.00752
        & -0.00354  & -0.00163  & -0.00049  & 0 \\
  0.3   & -0.06676  & -0.02167  & -0.00657  & -0.00191  & -0.00054
        & -0.00015  & -0.00004  & -0.00001  & 0 \\
  0.4   & -0.02942  & -0.00525  & -0.00087  & -0.00014  & -0.00002
        &  0.00000  &  0.00000  &  0.00000  & 0 \\
  0.5   & -0.00991  & -0.00088  & -0.00007  & -0.00001  &  0.00000
        &  0.00000  &  0.00000  &  0.00000  & 0 \\
  0. 6  & -0.00248  & -0.00010  &  0.00000  &  0.00000  &  0.00000
        &  0.00000  &  0.00000  &  0.00000  & 0 \\
  0.7   & -0.00045  & -0.00001  &  0.00000  &  0.00000  &  0.00000
        &  0.00000  &  0.00000  &  0.00000  & 0 \\
  0.8   & -0.00007  &  0.00000  &  0.00000  &  0.00000  &  0.00000
        &  0.00000  &  0.00000  &  0.00000  & 0 \\
  0.9   & -0.00002  &  0.00000  &  0.00000  &  0.00000  &  0.00000
        &  0.00000  &  0.00000  &  0.00000  & 0 \\
  1.0   & -0.00001  &  0.00000  &  0.00000  &  0.00000  &  0.00000
        &  0.00000  &  0.00000  &  0.00000  & 0 \\
  1.1   & -0.00001  &  0.00000  &  0.00000  &  0.00000  &  0.00000
        &  0.00000  &  0.00000  &  0.00000  & 0 \\
  1.2   &  0.00000  &  0.00000  &  0.00000  &  0.00000  &  0.00000
        &  0.00000  &  0.00000  &  0.00000  & 0 \\
  1.3   &  0.00005  &  0.00000  &  0.00000  &  0.00000  &  0.00000
        &  0.00000  &  0.00000  &  0.00000  & 0 \\
  1.4   &  0.00023  &  0.00002  &  0.00000  &  0.00000  &  0.00000
        &  0.00000  &  0.00000  &  0.00000  & 0 \\
  1.5   &  0.00040  &  0.00014  &  0.00002  &  0.00000  &  0.00000
        &  0.00000  &  0.00000  &  0.00000  & 0 \\
  1.6   & -0.00087  &  0.00066  &  0.00019  &  0.00004  &  0.00001
        &  0.00000  &  0.00000  &  0.00000  & 0 \\
  1.7   & -0.00818  &  0.00159  &  0.00125  &  0.00051  &  0.00017
        &  0.00005  &  0.00002  &  0.00000  & 0 \\
  1.8   & -0.02927  & -0.00149  &  0.00411  &  0.00367  &  0.00232
        &  0.00128  &  0.00066  &  0.00024  & 0 \\
  1.9   & -0.05656  & -0.02323  & -0.00366  &  0.00640  &  0.01050
        &  0.01123  &  0.01029  &  0.00673  & 0 \\ \hline    
\end{tabular}
\end{ruledtabular}
\end{table*}

\begin{table*}[htp]
\caption{\label{tab:residuall57lw}
Residual mass in lattice units at $\beta=5.7$, in Landau gauge 
for the L\"uscher-Weisz gauge action.} 
\begin{ruledtabular}    
\begin{tabular}{|c|rrrrrrrrr|} 
$M$ & $N_s=8$ & $N_s=12$ & $N_s=16$ & $N_s=20$ & $N_s=24$ & $N_s=28$ 
    & $N_s=32$ & $N_s=48$ & $N_s=\infty$  
      \vspace{0.05cm} \\ \hline \vspace{-0.3cm} \\  
  0.1   & -0.20427  & -0.18953  & -0.16862  & -0.14340  & -0.11716
        & -0.09262  & -0.07133  & -0.04383  & 0 \\
  0.2   & -0.21102  & -0.13029  & -0.07259  & -0.03775  & -0.01876
        & -0.00904  & -0.00425  & -0.00137  & 0 \\
  0.3   & -0.13548  & -0.04879  & -0.01566  & -0.00471  & -0.00136
        & -0.00038  & -0.00010  & -0.00002  & 0 \\
  0.4   & -0.06344  & -0.01227  & -0.00212  & -0.00034  & -0.00005
        & -0.00001  &  0.00000  &  0.00000  & 0 \\
  0.5   & -0.02250  & -0.00212  & -0.00018  & -0.00001  &  0.00000
        &  0.00000  &  0.00000  &  0.00000  & 0 \\
  0.6   & -0.00594  & -0.00024  & -0.00001  &  0.00000  &  0.00000
        &  0.00000  &  0.00000  &  0.00000  & 0 \\
  0.7   & -0.00114  & -0.00002  &  0.00000  &  0.00000  &  0.00000
        &  0.00000  &  0.00000  &  0.00000  & 0 \\
  0.8   & -0.00019  &  0.00000  &  0.00000  &  0.00000  &  0.00000
        &  0.00000  &  0.00000  &  0.00000  & 0 \\
  0.9   & -0.00006  &  0.00000  &  0.00000  &  0.00000  &  0.00000
        &  0.00000  &  0.00000  &  0.00000  & 0 \\
  1.0   & -0.00003  &  0.00000  &  0.00000  &  0.00000  &  0.00000
        &  0.00000  &  0.00000  &  0.00000  & 0 \\
  1.1   & -0.00002  &  0.00000  &  0.00000  &  0.00000  &  0.00000
        &  0.00000  &  0.00000  &  0.00000  & 0 \\
  1.2   &  0.00001  &  0.00000  &  0.00000  &  0.00000  &  0.00000
        &  0.00000  &  0.00000  &  0.00000  & 0 \\
  1.3   &  0.00038  &  0.00000  &  0.00000  &  0.00000  &  0.00000
        &  0.00000  &  0.00000  &  0.00000  & 0 \\
  1.4   &  0.00255  &  0.00011  &  0.00000  &  0.00000  &  0.00000
        &  0.00000  &  0.00000  &  0.00000  & 0 \\
  1.5   &  0.01012  &  0.00108  &  0.00010  &  0.00001  &  0.00000
        &  0.00000  &  0.00000  &  0.00000  & 0 \\
  1.6   &  0.02746  &  0.00649  &  0.00122  &  0.00021  &  0.00003
        &  0.00001  &  0.00000  &  0.00000  & 0 \\
  1.7   &  0.05182  &  0.02537  &  0.00924  &  0.00297  &  0.00090
        &  0.00026  &  0.00007  &  0.00002  & 0 \\
  1.8   &  0.05874  &  0.06132  &  0.04166  &  0.02399  &  0.01268
        &  0.00636  &  0.00309  &  0.00111  & 0 \\
  1.9   &  0.01626  &  0.05963  &  0.07909  &  0.08186  &  0.07499
        &  0.06379  &  0.05167  &  0.03349  & 0 \\ \hline
\end{tabular}
\end{ruledtabular}
\end{table*}

\begin{table*}[htp]
\caption{\label{tab:residuall26iw}
Residual mass in lattice units at $\beta=2.6$, in Landau gauge 
for the Iwasaki gauge action.} 
\begin{ruledtabular}    
\begin{tabular}{|c|rrrrrrrrr|} 
$M$ & $N_s=8$ & $N_s=12$ & $N_s=16$ & $N_s=20$ & $N_s=24$ & $N_s=28$ 
    & $N_s=32$ & $N_s=48$ & $N_s=\infty$  
      \vspace{0.05cm} \\ \hline \vspace{-0.3cm} \\  
  0.1   & -0.26041  & -0.25106  & -0.22837  & -0.19680  & -0.16214
        & -0.12890  & -0.09967  & -0.05976  & 0 \\
  0.2   & -0.28010  & -0.17816  & -0.10070  & -0.05280  & -0.02638
        & -0.01275  & -0.00602  & -0.00185  & 0 \\
  0.3   & -0.18434  & -0.06790  & -0.02204  & -0.00667  & -0.00193
        & -0.00054  & -0.00015  & -0.00002  & 0 \\
  0.4   & -0.08791  & -0.01732  & -0.00302  & -0.00049  & -0.00008
        & -0.00001  &  0.00000  &  0.00000  & 0 \\
  0.5   & -0.03170  & -0.00304  & -0.00026  & -0.00002  &  0.00000
        &  0.00000  &  0.00000  &  0.00000  & 0 \\
  0.6   & -0.00852  & -0.00035  & -0.00001  &  0.00000  &  0.00000
        &  0.00000  &  0.00000  &  0.00000  & 0 \\
  0.7   & -0.00167  & -0.00003  &  0.00000  &  0.00000  &  0.00000
        &  0.00000  &  0.00000  &  0.00000  & 0 \\
  0.8   & -0.00029  &  0.00000  &  0.00000  &  0.00000  &  0.00000
        &  0.00000  &  0.00000  &  0.00000  & 0 \\
  0.9   & -0.00008  &  0.00000  &  0.00000  &  0.00000  &  0.00000
        &  0.00000  &  0.00000  &  0.00000  & 0 \\
  1.0   & -0.00004  &  0.00000  &  0.00000  &  0.00000  &  0.00000
        &  0.00000  &  0.00000  &  0.00000  & 0 \\
  1.1   & -0.00003  &  0.00000  &  0.00000  &  0.00000  &  0.00000
        &  0.00000  &  0.00000  &  0.00000  & 0 \\
  1.2   &  0.00001  &  0.00000  &  0.00000  &  0.00000  &  0.00000
        &  0.00000  &  0.00000  &  0.00000  & 0 \\
  1.3   &  0.00053  &  0.00001  &  0.00000  &  0.00000  &  0.00000
        &  0.00000  &  0.00000  &  0.00000  & 0 \\
  1.4   &  0.00362  &  0.00015  &  0.00001  &  0.00000  &  0.00000
        &  0.00000  &  0.00000  &  0.00000  & 0 \\
  1.5   &  0.01465  &  0.00152  &  0.00013  &  0.00001  &  0.00000
        &  0.00000  &  0.00000  &  0.00000  & 0 \\
  1.6   &  0.04090  &  0.00928  &  0.00172  &  0.00029  &  0.00005
        &  0.00001  &  0.00000  &  0.00000  & 0 \\
  1.7   &  0.08062  &  0.03709  &  0.01323  &  0.00421  &  0.00126
        &  0.00036  &  0.00010  &  0.00002  & 0 \\
  1.8   &  0.10093  &  0.09276  &  0.06089  &  0.03453  &  0.01809
        &  0.00903  &  0.00436  &  0.00157  & 0 \\
  1.9   &  0.05046  &  0.10077  &  0.12163  &  0.12156  &  0.10945
        &  0.09220  &  0.07421  &  0.04790  & 0 \\ \hline
\end{tabular}
\end{ruledtabular}
\end{table*}

\begin{table*}[htp]
\caption{\label{tab:residuall104db}
Residual mass in lattice units at $\beta=1.04$, in Landau gauge 
for the DBW2 gauge action.} 
\begin{ruledtabular}    
\begin{tabular}{|c|rrrrrrrrr|} 
$M$ & $N_s=8$ & $N_s=12$ & $N_s=16$ & $N_s=20$ & $N_s=24$ & $N_s=28$ 
    & $N_s=32$ & $N_s=48$ & $N_s=\infty$  
      \vspace{0.05cm} \\ \hline \vspace{-0.3cm} \\  
  0.1   & -0.29645  & -0.28910  & -0.26373  & -0.22721  & -0.18696
        & -0.14842  & -0.11460  & -0.06565  & 0 \\
  0.2   & -0.32459  & -0.20788  & -0.11763  & -0.06168  & -0.03081
        & -0.01489  & -0.00703  & -0.00198  & 0 \\
  0.3   & -0.21621  & -0.08011  & -0.02607  & -0.00790  & -0.00229
        & -0.00065  & -0.00018  & -0.00003  & 0 \\
  0.4   & -0.10412  & -0.02066  & -0.00361  & -0.00059  & -0.00009
        & -0.00001  &  0.00000  &  0.00000  & 0 \\
  0.5   & -0.03784  & -0.00366  & -0.00031  & -0.00002  &  0.00000
        &  0.00000  &  0.00000  &  0.00000  & 0 \\
  0.6   & -0.01022  & -0.00043  & -0.00002  &  0.00000  &  0.00000
        &  0.00000  &  0.00000  &  0.00000  & 0 \\
  0.7   & -0.00200  & -0.00003  &  0.00000  &  0.00000  &  0.00000
        &  0.00000  &  0.00000  &  0.00000  & 0 \\
  0.8   & -0.00033  &  0.00000  &  0.00000  &  0.00000  &  0.00000
        &  0.00000  &  0.00000  &  0.00000  & 0 \\
  0.9   & -0.00009  &  0.00000  &  0.00000  &  0.00000  &  0.00000
        &  0.00000  &  0.00000  &  0.00000  & 0 \\
  1.0   & -0.00004  &  0.00000  &  0.00000  &  0.00000  &  0.00000
        &  0.00000  &  0.00000  &  0.00000  & 0 \\
  1.1   & -0.00003  &  0.00000  &  0.00000  &  0.00000  &  0.00000
        &  0.00000  &  0.00000  &  0.00000  & 0 \\
  1.2   &  0.00001  &  0.00000  &  0.00000  &  0.00000  &  0.00000
        &  0.00000  &  0.00000  &  0.00000  & 0 \\
  1.3   &  0.00049  &  0.00001  &  0.00000  &  0.00000  &  0.00000
        &  0.00000  &  0.00000  &  0.00000  & 0 \\
  1.4   &  0.00336  &  0.00014  &  0.00001  &  0.00000  &  0.00000
        &  0.00000  &  0.00000  &  0.00000  & 0 \\
  1.5   &  0.01371  &  0.00143  &  0.00013  &  0.00001  &  0.00000
        &  0.00000  &  0.00000  &  0.00000  & 0 \\
  1.6   &  0.03866  &  0.00889  &  0.00165  &  0.00028  &  0.00005
        &  0.00001  &  0.00000  &  0.00000  & 0 \\
  1.7   &  0.07672  &  0.03617  &  0.01301  &  0.00416  &  0.00125
        &  0.00036  &  0.00010  &  0.00002  & 0 \\
  1.8   &  0.09526  &  0.09152  &  0.06108  &  0.03493  &  0.01839
        &  0.00921  &  0.00446  &  0.00166  & 0 \\
  1.9   &  0.04437  &  0.09848  &  0.12253  &  0.12440  &  0.11311
        &  0.09591  &  0.07755  &  0.05076  & 0 \\ \hline
\end{tabular}
\end{ruledtabular}
\end{table*}

We can express $\Sigma_0$, the residual mass and the other quantities 
presented in this paper in the form
\begin{equation}
A + (1-\lambda) \, B ,
\end{equation}
where $A$ and $A+B$ provide the answers in Feynman and Landau gauge 
respectively, and $B$ is a number which remains the same when using fermion 
formulations rather diverse like domain-wall with an infinite extent of the 
fifth dimension, Wilson or overlap. Since, as we have noticed in the previous
Section, $B$ remains also the same whether the plaquette action or an improved 
gluon action is used, we refer for its results to the corresponding Tables 
in \cite{Capitani:2006kw}.

The values of $\Sigma_0$ which come out after our results for the tadpole 
and half-circle diagrams are added together are reported in Tables 
\ref{tab:sigma0lw}, \ref{tab:sigma0iw} and \ref{tab:sigma0db}.
The values of $a m_{res}^{(1)}$ which one obtains after the contribution of 
the tree-level residual mass is finally included are shown in Tables 
\ref{tab:residuall52lw} to \ref{tab:residuall52db} for the three improved gauge
actions considered. In the case of the DBW2 action, given that is the most used
in numerical simulations, we explicitly give the results for two choices of 
the bare coupling, $\beta=6.0$ and $\beta=5.2$, and in both Feynman and Landau 
gauges. This also helps to illustrate the fact that for the DBW2 action the 
violations of gauge invariance are here rather pronounced. 
The L\"uscher-Weisz and Iwasaki actions behave in this respect 
in a similar way, the main difference being that the residual mass is less 
small and the relative violations of gauge invariance are less large. 
For these two actions we only show here the Tables corresponding to $\beta=5.2$
in Landau gauge. Other choices of $\beta$ as well as results in Feynman gauge 
can be easily derived from the primary results provided for $\Sigma_0$.

In the plaquette case we had observed \cite{Capitani:2006kw} that when the 
one-loop corrections are taken into account, the residual mass, which is 
negative at tree level, changes sign and becomes positive. This also happens 
for the L\"uscher-Weisz and Iwasaki actions, and (except for very small $N_s$, 
or $M$ very close to 2) for the DBW2 action. We can check from the various 
Tables here provided that also for the improved gauge actions the residual mass
$a m_{res}^{(1)}$ happens to have the same sign of the critical mass of Wilson 
fermions (i.e., is positive) only for $M \gtrsim 1.2$, exactly as in the case 
of the plaquette action. This is the region of $M$ we are interested in. 
The value $M \simeq 1.2$ arises as a consequence of the additive 
renormalization undergone by $w_0$ (in this case, at one loop), and this shows 
that improved gauge actions do not behave too differently in this respect.

If we compare the results derived in this paper with the residual mass that 
was obtained using the plaquette gauge action \cite{Capitani:2006kw}, we can 
see that the use of improved gauge actions produces, for the same choice of 
$g_0$, a significant suppression of $a m_{res}^{(1)}$. 
A consistent picture of this effect can be gathered by 
looking at the numbers for $\beta=5.2$ in the Landau gauge for the various 
actions, contained in Tables \ref{tab:residual2l}, \ref{tab:residuall52lw}, 
\ref{tab:residuall52iw} and \ref{tab:residuall52db}. 
We can immediately see that the Iwasaki action gives a stronger suppression 
than the L\"uscher-Weisz action. Although not properly consistent at this 
order in $g_0$, we can also have a look at what happens in the case $N_s=16$ 
and $M=1.8$, which are typical values chosen in the numerical simulations. 
For these two actions, and for this choice, one-loop perturbative calculations 
give a residual mass which is respectively about two-thirds and one-third of 
the plaquette value. A very interesting outcome is that the DBW2 action is the 
most effective in generating a large suppression.
We can observe a general monotonic decrease of the 
residual mass when $c_1$ grows, however we can gather from Tables 
\ref{tab:residualf60db} to \ref{tab:residuall52db} that for the DBW2 action 
this suppression can at times go too far and for some choices of $N_s$ and $M$ 
we actually obtain (small) negative values for the residual mass. 

\begin{table}[htp]
\caption{\label{tab:comparison}
Results for the coefficient of $\bar g^2$ of the tadpole and half-circle 
diagrams, for $M=1.8$ and $N_s=16$ in Feynman gauge, and corresponding 
one-loop residual mass at $\beta = 5.2$. Plaquette numbers are taken from
\cite{Capitani:2006kw}. The tree-level residual mass is also shown.}
\begin{ruledtabular}    
\begin{tabular}{|l|rrrr|} 
   action & tadpole & half-circle & $am_{res}^{(0)}$ & $am_{res}^{(1)}$   
      \vspace{0.05cm} \\ \hline \vspace{-0.3cm} \\  
  plaquette        & -7.73201  &  0.36471  & -0.01013  &  0.06164  \\
  L\"uscher-Weisz  & -6.40240  &  0.71569  & -0.01013  &  0.04527  \\
  Iwasaki          & -4.72902  &  1.22505  & -0.01013  &  0.02400  \\
  DBW2             & -3.11540  &  1.79449  & -0.01013  &  0.00274  \\ \hline
\end{tabular}
\end{ruledtabular}
\end{table}

We should note at this point that the dramatic suppression of the residual mass
in the case of the DBW2 action is caused not only by the decreasing of the 
absolute value of the results for the tadpole (which are negative for $M > 1$),
but also by the increasing of the results for the half-circle diagram 
(which are positive). The numbers provided for $N_s=16$ and $M=1.8$ in Table 
\ref{tab:comparison} can summarize this behavior. We remind that $\Sigma_0$, 
which is the sum of these two diagrams, enters with a negative sign in the 
one-loop formula for $m_{res}$, Eq.~(\ref{eq:mres}). At the end, what happens 
is that the total contribution of the one-loop diagrams becomes smaller as 
$c_1$ grows, and since the (negative) tree-level residual mass remains always 
the same for all gauge actions, for the DBW2 action the compensations between 
the tree-level and one-loop diagrams are much larger. The final numbers for 
$m_{res}^{(1)}$ are then decisively smaller for this action, and sometimes 
they even overshoot and become negative.

Thus far we have only discussed comparisons made at the same value of the 
coupling among the various actions, and we have provided numbers for situations
in which $g_0^2$ is equal to 1 or near it. However, if we want to compare the 
results of the various actions at the same energy scale, we have to insert the 
appropriate values of $g_0$ belonging to each improved action for a given 
lattice spacing. The picture that comes out from such comparisons is then 
somewhat different. Let us take as an illustration the case of quenched
QCD at 2 GeV, a scale that Monte Carlo simulations teach us to be reached for 
$\beta=5.7$ when the L\"uscher-Weisz action is used, for $\beta=2.6$ when 
the Iwasaki action is used,and for $\beta=1.04$ when the DBW2 action is used. 
Note that here we maintain the definition of $\beta = 6/g_0^2$ also for the 
improved gauge actions, instead of $\beta' = 6 (1-8 c_1)/g_0^2$.
The corresponding results for $a m_{res}^{(1)}$ are collected in Tables 
\ref{tab:residuall57lw}, \ref{tab:residuall26iw} and \ref{tab:residuall104db}.
It is curious to notice from these Tables that the numbers obtained using 
the Iwasaki action are rather close to those obtained using the DBW2 action,
and surprisingly they lie in general slightly above the plaquette values 
(which we report in Table \ref{tab:residual1l} in the Appendix). 

We have however to stress at this point that for the quenched DBW2 case at 
2 GeV the gauge coupling assumes the value $g_0^2 = 5.77$, which appears 
to be somewhat too large to constitute a reasonable perturbative expansion 
coefficient. Moreover, such values of the coupling are derived solely from 
numerical simulations, and they then contain informations of nonperturbative 
nature as well, so that a mismatch can arise when one only takes into account 
the results of the one-loop diagrams calculated at these values of the 
coupling, that is when $c_1$ is not small.

Notice that the L\"uscher-Weisz action gives in the above case the best results
in terms of suppression, and indeed $g_0^2$ is still not too far from 1 for 
this action even at the scale of 2 GeV.

Although it is clear that one-loop perturbation theory encounters here some of 
its limitations and cannot provide the whole story, the small numbers for the 
residual mass that we can see in the Tables for the DBW2 action when $g_0^2$ 
is near 1 may qualitatively confirm the knowledge about the effects of improved
gauge actions which has been gathered from numerical simulations in the past 
years. Since one-loop perturbation theory is in this case not fully adequate, 
2-loop (and higher) corrections should be considered (at least for the Iwasaki 
and DBW2 actions) if one is interested in getting more reliable numbers. 
It looks like the two-loop contributions become indeed more and more important 
as $c_1$ grows, because either there are ever larger compensations between 
one-loop and tree-level diagrams when $g_0^2 \sim 1$, or values of the coupling
that are too large appear in the other cases.  

The difference between the Landau and Feynman gauge results, which in absolute 
numbers is the same for all the actions considered (and for the plaquette 
action), in the case of the DBW2 gauge action, where the numbers are rather 
small, comes out of the same order as the results. The violations of gauge 
invariance are then proportionally more significant here.

All numbers that we have obtained are valid for the quenched as well as the 
unquenched case, because at one loop internal quark loops can never appear 
in the diagrams that enter in this as well as in the other calculations 
presented in this paper (see Figure \ref{fig:diagrams}). Since the residual 
mass has an obvious dependence on the coupling $g_0$ and also depends 
explicitly on the value of the lattice spacing $a$ (it is in fact given in 
lattice units), it is different for quenched and unquenched simulations 
made at the same lattice spacing, given that $a$ and $g_0$ are related in 
a different way. 

Many other general features that we have encountered in the case of the 
calculations made with the plaquette action are also present for improved gauge
actions. Thus, the deviations from the case of exact chiral symmetry are rather
large when $N_s$ is very small or $M$ is close to 0.1 or 1.9, and, as we have
seen, the residual mass is positive (with some exceptions) only when 
$M \gtrsim 1.2$. The rate of decay in $N_s$ at fixed $M$, which is connected 
to the value of the mobility edge $\lambda_c$ \cite{Golterman:2003qe,Golterman:2004cy,Golterman:2005fe,Christ:2005xh,Svetitsky:2005qa},
\begin{equation}
m_{res} \sim R^4_e \, \rho_e (\lambda_c) \, \frac{\exp{(-\lambda_c N_s)}}{N_s}
+ R^4_l \, \rho_l (0) \, \frac{1}{N_s} 
\label{eq:transmat}
\end{equation}
(where $\rho$ is the density per unit spacetime volume of the eigenvalues 
of the logarithm of the transfer matrix, and $l$ and $e$ stand for localized
and extended modes with average size $R$ respectively), increases when $M$ 
grows past 1.2 and reaches a maximum before it starts to decrease, approaching
then zero near the border $M=2$. Thus, these decays in $N_s$ still slow down 
when one approaches the borders of allowed values of $M$ ($M=0$ and $M=2$).
This can be related to the decrease of the mobility edge $\lambda_c$ towards 
zero in these extreme regions of $M$ 
\cite{Golterman:2003qe,Golterman:2004cy,Golterman:2005fe}, which signals the 
onset of the Aoki phase. 
Furthermore, the phenomena happening near the borders that were discussed in 
\cite{Capitani:2006kw} are also present here. Near $M=2$ one can indeed notice 
that the residual mass initially grows with $N_s$ (at fixed $M$) before the 
exponential decay finally sets in. This can be seen for example in Table 
\ref{tab:residuall52lw}, where already for $M=1.9$ the residual mass at 
$N_s=12$ is larger than at $N_s=8$, and at $N_s=16$ is even larger. 
The only difference with the plaquette case is that the numbers involved are 
here smaller, and sometimes they can even become negative, as in the DBW2 case,
however as we have already remarked in this case a two-loop calculation would  
likely be more appropriate.

For more discussion about these phenomena we refer again to Ref.
\cite{Capitani:2006kw}.

\section{Bilinear differences and a power-divergent mixing}
\label{sec:vad1}

Along the same lines as in \cite{Capitani:2006kw} we have also computed the 
quantity $\Delta = Z_V - Z_A = -(Z_S - Z_P)/2$, which only becomes zero in
the limit of infinite $N_s$, that is when chiral symmetry is fully restored. 
Since the vector and axial-vector currents renormalize differently when chiral 
symmetry is broken, this quantity can provide another estimate of 
chirality-breaking effects.

As we can gather from the results in Feynman gauge presented in Tables 
\ref{tab:valw} to \ref{tab:vadb}, the amount of chirality breaking connected 
to $\Delta$ follows a pattern similar to the one that we have encountered in 
the case of the residual mass, that is $\Delta$ is relatively large for small 
$N_s$ and $|1-M|$ close to 1, and decreases when $N_s$ grows or when $|1-M|$ 
tends toward zero. As we have already remarked, the part proportional to 
$1-\lambda$ of $\Delta$ is the same as for the plaquette action, which 
implies a violation of gauge invariance.

We can however observe here that, if one carries out comparisons at the same 
value of the gauge coupling, the use of improved gauge actions produces only 
small changes in the results for $\Delta$ compared to the plaquette action, 
and thus does not further suppress this quantity. When things are instead 
reported to the same energy scale, it is clear that the much increased value 
of the coupling of the DBW2 action enhances the corresponding results, and 
of a large factor. This is another reason to cast doubts about the adequacy 
of one-loop calculations in this case, as we remarked for the residual mass.
In any case for the 
plaquette $\Delta$ was already much smaller (at given $M$ and $N_s$) than 
$m_{res}$, being probably of a higher order in $m_{res}$ as it has been 
suggested for four-fermion operators in \cite{Aoki:2004ht,Christ:2005xh}.
Looking again at the nonperturbative choice $M=1.8$ and $N_s=16$, the numbers 
that we have obtained would imply that when one increases $c_1$ from the 
plaquette to the DBW2 action the change amounts to only a few percent at most. 

Thus, somewhat surprisingly, in the case of $\Delta$ improved gauge actions 
do not seem to make things better (at the same $g_0$) compared with the 
plaquette action .
If for this quantity a reduction due to the use of improved gauge actions is
evinced from numerical simulations, it should probably be traced entirely to 
nonperturbative effects.

We have also considered the power-divergent mixing (which is nonzero only 
in the case in which chiral symmetry is broken) of 
\cite{Capitani:2000aq,Capitani:2005vb} 
\begin{equation}
O_{d_1} = \bar q \gamma_{[4} \gamma_5 D_{1]} q ,
\end{equation}
which is an operator related to the distribution of the (chiral even) 
transverse spin of quarks inside hadrons. Its mixing with an operator of 
lower dimension can be written as
\begin{equation}
c_{mix} \cdot \frac{i}{a} \, \bar q \sigma_{41} \gamma_5 q .
\label{eq:g2mixing}
\end{equation}
For finite $N_s$ the results for $c_{mix}$ in Feynman gauge are reported in 
Tables \ref{tab:d1mixlw} to \ref{tab:d1mixdb}, where we can see that in general
this mixing is almost negligible. Were this not the case, the removal of these 
lattice artifacts in Monte Carlo simulations of domain-wall QCD would become 
quite challenging.

Thus, as in the case of $\Delta$, these chiral violations seem to be of higher 
order in $m_{res}$, and in general rather small. However here we can see that,
at variance with $\Delta$, improved gauge actions do suppress the amount of 
mixing (at the same $g_0$), with the DBW2 action producing a suppression of 
roughly one order of magnitude. This looks similar to what happens in the case 
of $m_{res}$.
We can also notice that when $c_1$ increases from the Iwasaki to 
the DBW2 action the results happen to change sign in the region around 
$M=1.8-1.9$, perhaps again indicating that the suppression goes too far and 
two-loop corrections are to be taken into account here.

We would like now to understand the different effects relative to $c_{mix}$ 
and $\Delta$ that we have obtained in this Section. We can begin by noticing 
that, unlike the residual mass, there is no tree-level contribution to $\Delta$
and $c_{mix}$. Moreover, no tadpole enters in the one-loop calculations of 
these quantities, and certainly there cannot be any tadpole dominance here. 
Furthermore, still at variance with the residual mass, the half-circle diagram 
is also absent. In fact, in the case of $\Delta$ this diagrams cancels in the 
difference of $V$ and $A$, while for $c_{mix}$ it is not present from the 
start. Only the vertex diagram contributes then to the calculation of $\Delta$,
but in the case of $c_{mix}$ the sail diagrams (see Figure \ref{fig:diagrams})
are present as well, because the operator contains a gauge field coming from 
the covariant derivative. This could provide the explanation for the different 
behaviors of $\Delta$ and $c_{mix}$ when improved gauge actions are turned on. 
If one looks at the perturbative results which were obtained in the past 
employing improved gauge actions with Wilson 
\cite{Aoki:1998ar,Taniguchi:1998zb,Aoki:2000ps} and overlap fermions 
\cite{Horsley:2004mx,Horsley:2005jk,Horsley:2005dx,Ioannou:2005uu,Ioannou:2006ds}, 
and also with domain-wall fermions at infinite $N_s$ \cite{Aoki:2002iq},
one can notice that the numbers for the vertex diagrams do not change very 
much when going from the plaquette to improved gauge actions. For example, in 
the case of Wilson fermions the lattice finite constant of the vertex diagram 
of the vector current is 8.765394 for the plaquette action and 7.050662
for the DBW2 action. For the axial-vector current these values are 3.943879 
and 5.555827 respectively. It is in general the tadpole and half-circle 
diagrams that end up generating the largest changes in the renormalization of 
these currents, but as we have remarked these diagrams do not contribute 
to $\Delta$ and $c_{mix}$.

These past results for the vertex diagrams make plausible our result that 
$\Delta$ is essentially not changed (at the same $g_0$) by the use of improved 
gauge actions
in the case of domain-wall fermions at finite $N_s$. They however refer to 
multiplicative renormalization factors of simple bilinear operators, and things
could behave differently in the case of the power-divergent mixing coefficient 
$c_{mix}$. In looking for an explanation for the behavior of $c_{mix}$, we 
have then computed this quantity using improved gauge actions also in the case 
of standard Wilson fermions (for domain-wall at infinite $N_s$ and overlap 
this coefficient is instead zero). To our knowledge these numbers had not been 
computed so far. The results of our Wilson calculations for $c_{mix}$ are 
collected in Table \ref{tab:wilsoncmix}. We have also computed it with 
Sheikholeslami-Wohlert fermionic improvement, and we give its numbers for 
$c_{sw}=1$ and $c_{sw}=2$, which are sufficient, since there are at most two 
improved vertices in the diagrams, to reconstruct $c_{mix}$ also for any value 
of $c_{sw}$. It can then be seen that the combined use of fermionic improvement
(with $c_{sw}=1$) and improved gauge actions (especially the DBW2 action) 
suppresses this mixing considerably (a factor 11.25 in this case). This effect 
also recalls to mind what happens when UV-filtering and fermionic improvement 
are combined together \cite{Capitani:2006ni}. The numbers in Table 
\ref{tab:wilsoncmix} seem to make plausible the behavior of $c_{mix}$ with 
improved gauge actions that we have found for domain-wall fermions at finite 
$N_s$, and in some sense the domain-wall formulation can be thought as a 
collection of many Wilson fermions with the appropriate damping factors.
On closer inspection, it turns out that in both formulations (Wilson and 
domain-wall) the bulk of this decrease is produced by the sail diagrams 
(present here because of the covariant derivative in the operator), which 
then appear to be the prime responsible for the different behavior of 
$\Delta$ and $c_{mix}$.

We can also take out as a lesson from the above considerations that the effect 
of improved gauge actions in the case of $\Delta$ and $c_{mix}$ remains
substantially the same in the case of Wilson and domain-wall fermions. 

Finally, we remark that while it is true that the combination of improved gauge
actions and fermionic improvement strongly suppresses $c_{mix}$ also in the 
case of the Wilson action (again, if one means that the comparisons are made 
at the same coupling), we should not overlook that (apart from extreme 
choices of $N_s$ and $M$) the numbers that come out from the use of domain-wall
fermions are much smaller than the ones of Wilson fermions.

\begin{table}[htp]
\caption{\label{tab:wilsoncmix}
Results for the coefficient of $\bar g^2$ of $c_{mix}$ for standard Wilson
fermions (including fermionic improvement).}
\begin{ruledtabular}    
\begin{tabular}{|l|rrr|} 
   action & $c_{sw}=0$ &  $c_{sw}=1$ & $c_{sw}=2$    
      \vspace{0.05cm} \\ \hline \vspace{-0.3cm} \\  
  plaquette        & 16.243762  &  8.798732  &  0.174259  \\
  L\"uscher-Weisz  & 13.517293  &  6.887859  & -0.749841  \\
  Iwasaki          &  9.461302  &  4.216070  & -1.756792  \\
  DBW2             &  4.688662  &  1.443948  & -2.158332  \\ \hline
\end{tabular}
\end{ruledtabular}
\end{table}

\begin{table*}[htp]
\caption{\label{tab:valw}
Coefficient of $\bar g^2$ for the quantity $\Delta = Z_V - Z_A = -(Z_S-Z_P)/2$,
in Feynman gauge for the L\"uscher-Weisz gauge action.}
\begin{ruledtabular}    
\begin{tabular}{|c|rrrrrrrrr|} 
$M$ & $N_s=8$ & $N_s=12$ & $N_s=16$ & $N_s=20$ & $N_s=24$ & $N_s=28$ 
    & $N_s=32$ & $N_s=48$ & $N_s=\infty$  
      \vspace{0.05cm} \\ \hline \vspace{-0.3cm} \\  
  0.1   &  6.6748  &  4.3283  &  2.6444  &  1.5535  &  0.8823
        &  0.4859  &  0.2605  &  0.0178  & 0 \\
  0.2   &  2.2664  &  0.7107  &  0.1928  &  0.0474  &  0.0110
        &  0.0024  &  0.0005  &  0.0000  & 0 \\
  0.3   &  0.5778  &  0.0662  &  0.0064  &  0.0006  &  0.0000
        &  0.0000  &  0.0000  &  0.0000  & 0 \\
  0.4   &  0.0976  &  0.0034  &  0.0001  &  0.0000  &  0.0000
        &  0.0000  &  0.0000  &  0.0000  & 0 \\
  0.5   &  0.0104  &  0.0001  &  0.0000  &  0.0000  &  0.0000
        &  0.0000  &  0.0000  &  0.0000  & 0 \\
  0.6   &  0.0006  &  0.0000  &  0.0000  &  0.0000  &  0.0000
        &  0.0000  &  0.0000  &  0.0000  & 0 \\
  0.7   &  0.0000  &  0.0000  &  0.0000  &  0.0000  &  0.0000
        &  0.0000  &  0.0000  &  0.0000  & 0 \\
  0.8   &  0.0000  &  0.0000  &  0.0000  &  0.0000  &  0.0000
        &  0.0000  &  0.0000  &  0.0000  & 0 \\
  0.9   &  0.0000  &  0.0000  &  0.0000  &  0.0000  &  0.0000
        &  0.0000  &  0.0000  &  0.0000  & 0 \\
  1.0   &  0.0000  &  0.0000  &  0.0000  &  0.0000  &  0.0000
        &  0.0000  &  0.0000  &  0.0000  & 0 \\
  1.1   &  0.0000  &  0.0000  &  0.0000  &  0.0000  &  0.0000
        &  0.0000  &  0.0000  &  0.0000  & 0 \\
  1.2   &  0.0000  &  0.0000  &  0.0000  &  0.0000  &  0.0000
        &  0.0000  &  0.0000  &  0.0000  & 0 \\
  1.3   &  0.0000  &  0.0000  &  0.0000  &  0.0000  &  0.0000
        &  0.0000  &  0.0000  &  0.0000  & 0 \\
  1.4   &  0.0005  &  0.0000  &  0.0000  &  0.0000  &  0.0000
        &  0.0000  &  0.0000  &  0.0000  & 0 \\
  1.5   &  0.0083  &  0.0001  &  0.0000  &  0.0000  &  0.0000
        &  0.0000  &  0.0000  &  0.0000  & 0 \\
  1.6   &  0.0698  &  0.0029  &  0.0001  &  0.0000  &  0.0000
        &  0.0000  &  0.0000  &  0.0000  & 0 \\
  1.7   &  0.3112  &  0.0500  &  0.0056  &  0.0005  &  0.0000
        &  0.0000  &  0.0000  &  0.0000  & 0 \\
  1.8   &  0.8168  &  0.3640  &  0.1283  &  0.0373  &  0.0095
        &  0.0022  &  0.0005  &  0.0000  & 0 \\
  1.9   &  2.1037  &  1.4869  &  1.0150  &  0.6819  &  0.4453
        &  0.2795  &  0.1679  &  0.0152  & 0 \\ \hline
\end{tabular}
\end{ruledtabular}
\end{table*}

\begin{table*}[htp]
\caption{\label{tab:vaiw}
Coefficient of $\bar g^2$ for the quantity $\Delta = Z_V - Z_A = -(Z_S-Z_P)/2$,
in Feynman gauge for the Iwasaki gauge action.}
\begin{ruledtabular}    
\begin{tabular}{|c|rrrrrrrrr|} 
$M$ & $N_s=8$ & $N_s=12$ & $N_s=16$ & $N_s=20$ & $N_s=24$ & $N_s=28$ 
    & $N_s=32$ & $N_s=48$ & $N_s=\infty$  
      \vspace{0.05cm} \\ \hline \vspace{-0.3cm} \\  
  0.1   &  6.2243  &  4.1314  &  2.5590  &  1.5166  &  0.8664
        &  0.4791  &  0.2575  &  0.0177  & 0 \\
  0.2   &  2.1620  &  0.6927  &  0.1897  &  0.0469  &  0.0109
        &  0.0024  &  0.0005  &  0.0000  & 0 \\
  0.3   &  0.5578  &  0.0650  &  0.0063  &  0.0006  &  0.0000
        &  0.0000  &  0.0000  &  0.0000  & 0 \\
  0.4   &  0.0945  &  0.0033  &  0.0001  &  0.0000  &  0.0000
        &  0.0000  &  0.0000  &  0.0000  & 0 \\
  0.5   &  0.0100  &  0.0001  &  0.0000  &  0.0000  &  0.0000
        &  0.0000  &  0.0000  &  0.0000  & 0 \\
  0.6   &  0.0006  &  0.0000  &  0.0000  &  0.0000  &  0.0000
        &  0.0000  &  0.0000  &  0.0000  & 0 \\
  0.7   &  0.0000  &  0.0000  &  0.0000  &  0.0000  &  0.0000
        &  0.0000  &  0.0000  &  0.0000  & 0 \\
  0.8   &  0.0000  &  0.0000  &  0.0000  &  0.0000  &  0.0000
        &  0.0000  &  0.0000  &  0.0000  & 0 \\
  0.9   &  0.0000  &  0.0000  &  0.0000  &  0.0000  &  0.0000
        &  0.0000  &  0.0000  &  0.0000  & 0 \\
  1.0   &  0.0000  &  0.0000  &  0.0000  &  0.0000  &  0.0000
        &  0.0000  &  0.0000  &  0.0000  & 0 \\
  1.1   &  0.0000  &  0.0000  &  0.0000  &  0.0000  &  0.0000
        &  0.0000  &  0.0000  &  0.0000  & 0 \\
  1.2   &  0.0000  &  0.0000  &  0.0000  &  0.0000  &  0.0000
        &  0.0000  &  0.0000  &  0.0000  & 0 \\
  1.3   &  0.0000  &  0.0000  &  0.0000  &  0.0000  &  0.0000
        &  0.0000  &  0.0000  &  0.0000  & 0 \\
  1.4   &  0.0005  &  0.0000  &  0.0000  &  0.0000  &  0.0000
        &  0.0000  &  0.0000  &  0.0000  & 0 \\
  1.5   &  0.0082  &  0.0001  &  0.0000  &  0.0000  &  0.0000
        &  0.0000  &  0.0000  &  0.0000  & 0 \\
  1.6   &  0.0694  &  0.0029  &  0.0001  &  0.0000  &  0.0000
        &  0.0000  &  0.0000  &  0.0000  & 0 \\
  1.7   &  0.3073  &  0.0497  &  0.0056  &  0.0005  &  0.0000
        &  0.0000  &  0.0000  &  0.0000  & 0 \\
  1.8   &  0.7891  &  0.3592  &  0.1275  &  0.0372  &  0.0095
        &  0.0022  &  0.0005  &  0.0000  & 0 \\
  1.9   &  1.9492  &  1.4179  &  0.9846  &  0.6686  &  0.4396
        &  0.2770  &  0.1668  &  0.0152  & 0 \\ \hline
\end{tabular}
\end{ruledtabular}
\end{table*}

\begin{table*}[htp]
\caption{\label{tab:vadb}
Coefficient of $\bar g^2$ for the quantity $\Delta = Z_V - Z_A = -(Z_S-Z_P)/2$,
in Feynman gauge for the DBW2 gauge action.}
\begin{ruledtabular}    
\begin{tabular}{|c|rrrrrrrrr|} 
$M$ & $N_s=8$ & $N_s=12$ & $N_s=16$ & $N_s=20$ & $N_s=24$ & $N_s=28$ 
    & $N_s=32$ & $N_s=48$ & $N_s=\infty$  
      \vspace{0.05cm} \\ \hline \vspace{-0.3cm} \\  
  0.1   &  5.4947  &  3.8063  &  2.4168  &  1.4548  &  0.8396
        &  0.4675  &  0.2525  &  0.0175  & 0 \\
  0.2   &  1.9894  &  0.6623  &  0.1845  &  0.0460  &  0.0108
        &  0.0024  &  0.0005  &  0.0000  & 0 \\
  0.3   &  0.5250  &  0.0629  &  0.0062  &  0.0005  &  0.0000
        &  0.0000  &  0.0000  &  0.0000  & 0 \\
  0.4   &  0.0896  &  0.0032  &  0.0001  &  0.0000  &  0.0000
        &  0.0000  &  0.0000  &  0.0000  & 0 \\
  0.5   &  0.0094  &  0.0001  &  0.0000  &  0.0000  &  0.0000
        &  0.0000  &  0.0000  &  0.0000  & 0 \\
  0.6   &  0.0005  &  0.0000  &  0.0000  &  0.0000  &  0.0000
        &  0.0000  &  0.0000  &  0.0000  & 0 \\
  0.7   &  0.0000  &  0.0000  &  0.0000  &  0.0000  &  0.0000
        &  0.0000  &  0.0000  &  0.0000  & 0 \\
  0.8   &  0.0000  &  0.0000  &  0.0000  &  0.0000  &  0.0000
        &  0.0000  &  0.0000  &  0.0000  & 0 \\
  0.9   &  0.0000  &  0.0000  &  0.0000  &  0.0000  &  0.0000
        &  0.0000  &  0.0000  &  0.0000  & 0 \\
  1.0   &  0.0000  &  0.0000  &  0.0000  &  0.0000  &  0.0000
        &  0.0000  &  0.0000  &  0.0000  & 0 \\
  1.1   &  0.0000  &  0.0000  &  0.0000  &  0.0000  &  0.0000
        &  0.0000  &  0.0000  &  0.0000  & 0 \\
  1.2   &  0.0000  &  0.0000  &  0.0000  &  0.0000  &  0.0000
        &  0.0000  &  0.0000  &  0.0000  & 0 \\
  1.3   &  0.0000  &  0.0000  &  0.0000  &  0.0000  &  0.0000
        &  0.0000  &  0.0000  &  0.0000  & 0 \\
  1.4   &  0.0005  &  0.0000  &  0.0000  &  0.0000  &  0.0000
        &  0.0000  &  0.0000  &  0.0000  & 0 \\
  1.5   &  0.0082  &  0.0001  &  0.0000  &  0.0000  &  0.0000
        &  0.0000  &  0.0000  &  0.0000  & 0 \\
  1.6   &  0.0688  &  0.0029  &  0.0001  &  0.0000  &  0.0000
        &  0.0000  &  0.0000  &  0.0000  & 0 \\
  1.7   &  0.3021  &  0.0494  &  0.0055  &  0.0005  &  0.0000
        &  0.0000  &  0.0000  &  0.0000  & 0 \\
  1.8   &  0.7516  &  0.3521  &  0.1262  &  0.0369  &  0.0095
        &  0.0022  &  0.0005  &  0.0000  & 0 \\
  1.9   &  1.7396  &  1.3168  &  0.9384  &  0.6480  &  0.4305
        &  0.2730  &  0.1651  &  0.0151  & 0 \\ \hline
\end{tabular}
\end{ruledtabular}
\end{table*}

\begin{table*}[htp]
\caption{\label{tab:d1mixlw}
Coefficient of $\bar g^2$ for the coefficient of the power-divergent mixing 
of $O_{d_1}$, $c_{mix}$ in Eq.~(\ref{eq:g2mixing}), in Feynman gauge
for the L\"uscher-Weisz gauge action.}
\begin{ruledtabular}    
\begin{tabular}{|c|rrrrrrrrr|} 
$M$ & $N_s=8$ & $N_s=12$ & $N_s=16$ & $N_s=20$ & $N_s=24$ & $N_s=28$ 
    & $N_s=32$ & $N_s=48$ & $N_s=\infty$  
      \vspace{0.05cm} \\ \hline \vspace{-0.3cm} \\  
  0.1   &-12.2811  & -7.3668  & -4.6923  & -3.0533  & -2.0014
        & -1.3145  & -0.8635  & -0.1604  & 0 \\
  0.2   & -5.0766  & -2.1043  & -0.8697  & -0.3578  & -0.1471
        & -0.0605  & -0.0249  & -0.0007  & 0 \\
  0.3   & -2.1594  & -0.5314  & -0.1293  & -0.0315  & -0.0077
        & -0.0019  & -0.0005  &  0.0000  & 0 \\
  0.4   & -0.8140  & -0.1104  & -0.0148  & -0.0020  & -0.0003
        &  0.0000  &  0.0000  &  0.0000  & 0 \\
  0.5   & -0.2675  & -0.0186  & -0.0012  & -0.0001  &  0.0000
        &  0.0000  &  0.0000  &  0.0000  & 0 \\
  0.6   & -0.0768  & -0.0025  & -0.0001  &  0.0000  &  0.0000
        &  0.0000  &  0.0000  &  0.0000  & 0 \\
  0.7   & -0.0197  & -0.0003  &  0.0000  &  0.0000  &  0.0000
        &  0.0000  &  0.0000  &  0.0000  & 0 \\
  0.8   & -0.0050  & -0.0001  &  0.0000  &  0.0000  &  0.0000
        &  0.0000  &  0.0000  &  0.0000  & 0 \\
  0.9   & -0.0013  &  0.0000  &  0.0000  &  0.0000  &  0.0000
        &  0.0000  &  0.0000  &  0.0000  & 0 \\
  1.0   &  0.0000  &  0.0000  &  0.0000  &  0.0000  &  0.0000
        &  0.0000  &  0.0000  &  0.0000  & 0 \\
  1.1   &  0.0006  &  0.0000  &  0.0000  &  0.0000  &  0.0000
        &  0.0000  &  0.0000  &  0.0000  & 0 \\
  1.2   &  0.0011  &  0.0000  &  0.0000  &  0.0000  &  0.0000
        &  0.0000  &  0.0000  &  0.0000  & 0 \\
  1.3   &  0.0024  &  0.0000  &  0.0000  &  0.0000  &  0.0000
        &  0.0000  &  0.0000  &  0.0000  & 0 \\
  1.4   &  0.0097  &  0.0002  &  0.0000  &  0.0000  &  0.0000
        &  0.0000  &  0.0000  &  0.0000  & 0 \\
  1.5   &  0.0409  &  0.0021  &  0.0001  &  0.0000  &  0.0000
        &  0.0000  &  0.0000  &  0.0000  & 0 \\
  1.6   &  0.1467  &  0.0142  &  0.0015  &  0.0002  &  0.0000
        &  0.0000  &  0.0000  &  0.0000  & 0 \\
  1.7   &  0.4250  &  0.0767  &  0.0142  &  0.0028  &  0.0006
        &  0.0001  &  0.0000  &  0.0000  & 0 \\
  1.8   &  0.9787  &  0.3200  &  0.1056  &  0.0345  &  0.0113
        &  0.0038  &  0.0013  &  0.0000  & 0 \\
  1.9   &  1.8728  &  0.9467  &  0.5303  &  0.3068  &  0.1790
        &  0.1043  &  0.0605  &  0.0064  & 0 \\ \hline
\end{tabular}
\end{ruledtabular}
\end{table*}

\begin{table*}[htp]
\caption{\label{tab:d1mixiw}
Coefficient of $\bar g^2$ for the coefficient of the power-divergent mixing 
of $O_{d_1}$, $c_{mix}$ in Eq.~(\ref{eq:g2mixing}), in Feynman gauge
for the Iwasaki gauge action.}
\begin{ruledtabular}    
\begin{tabular}{|c|rrrrrrrrr|} 
$M$ & $N_s=8$ & $N_s=12$ & $N_s=16$ & $N_s=20$ & $N_s=24$ & $N_s=28$ 
    & $N_s=32$ & $N_s=48$ & $N_s=\infty$  
      \vspace{0.05cm} \\ \hline \vspace{-0.3cm} \\  
  0.1   & -9.4574  & -5.7117  & -3.6538  & -2.3844  & -1.5659
        & -1.0297  & -0.6769  & -0.1259  & 0 \\
  0.2   & -3.9356  & -1.6453  & -0.6821  & -0.2810  & -0.1156
        & -0.0476  & -0.0196  & -0.0006  & 0 \\
  0.3   & -1.6731  & -0.4139  & -0.1010  & -0.0247  & -0.0060
        & -0.0015  & -0.0004  &  0.0000  & 0 \\
  0.4   & -0.6240  & -0.0853  & -0.0115  & -0.0015  & -0.0002
        &  0.0000  &  0.0000  &  0.0000  & 0 \\
  0.5   & -0.2020  & -0.0142  & -0.0010  & -0.0001  &  0.0000
        &  0.0000  &  0.0000  &  0.0000  & 0 \\
  0.6   & -0.0570  & -0.0019  & -0.0001  &  0.0000  &  0.0000
        &  0.0000  &  0.0000  &  0.0000  & 0 \\
  0.7   & -0.0143  & -0.0002  &  0.0000  &  0.0000  &  0.0000
        &  0.0000  &  0.0000  &  0.0000  & 0 \\
  0.8   & -0.0035  &  0.0000  &  0.0000  &  0.0000  &  0.0000
        &  0.0000  &  0.0000  &  0.0000  & 0 \\
  0.9   & -0.0009  &  0.0000  &  0.0000  &  0.0000  &  0.0000
        &  0.0000  &  0.0000  &  0.0000  & 0 \\
  1.0   &  0.0000  &  0.0000  &  0.0000  &  0.0000  &  0.0000
        &  0.0000  &  0.0000  &  0.0000  & 0 \\
  1.1   &  0.0004  &  0.0000  &  0.0000  &  0.0000  &  0.0000
        &  0.0000  &  0.0000  &  0.0000  & 0 \\
  1.2   &  0.0007  &  0.0000  &  0.0000  &  0.0000  &  0.0000
        &  0.0000  &  0.0000  &  0.0000  & 0 \\
  1.3   &  0.0015  &  0.0000  &  0.0000  &  0.0000  &  0.0000
        &  0.0000  &  0.0000  &  0.0000  & 0 \\
  1.4   &  0.0054  &  0.0001  &  0.0000  &  0.0000  &  0.0000
        &  0.0000  &  0.0000  &  0.0000  & 0 \\  
  1.5   &  0.0218  &  0.0009  &  0.0000  &  0.0000  &  0.0000
        &  0.0000  &  0.0000  &  0.0000  & 0 \\
  1.6   &  0.0773  &  0.0053  &  0.0004  &  0.0000  &  0.0000
        &  0.0000  &  0.0000  &  0.0000  & 0 \\
  1.7   &  0.2172  &  0.0273  &  0.0024  & -0.0001  & -0.0001
        &  0.0000  &  0.0000  &  0.0000  & 0 \\
  1.8   &  0.4328  &  0.1029  &  0.0172  & -0.0016  & -0.0035
        & -0.0023  & -0.0012  &  0.0000  & 0 \\
  1.9   &  0.4010  &  0.0932  & -0.0035  & -0.0366  & -0.0444
        & -0.0417  & -0.0352  & -0.0113  & 0 \\ \hline
\end{tabular}
\end{ruledtabular}
\end{table*}

\begin{table*}[htp]
\caption{\label{tab:d1mixdb}
Coefficient of $\bar g^2$ for the coefficient of the power-divergent mixing 
of $O_{d_1}$, $c_{mix}$ in Eq.~(\ref{eq:g2mixing}), in Feynman gauge
for the DBW2 gauge action.}
\begin{ruledtabular}    
\begin{tabular}{|c|rrrrrrrrr|} 
$M$ & $N_s=8$ & $N_s=12$ & $N_s=16$ & $N_s=20$ & $N_s=24$ & $N_s=28$ 
    & $N_s=32$ & $N_s=48$ & $N_s=\infty$  
      \vspace{0.05cm} \\ \hline \vspace{-0.3cm} \\  
  0.1   & -5.8666  & -3.5954  & -2.3230  & -1.5265  & -1.0072
        & -0.6642  & -0.4374  & -0.0815  & 0 \\
  0.2   & -2.4673  & -1.0511  & -0.4388  & -0.1812  & -0.0747
        & -0.0308  & -0.0128  & -0.0004  & 0 \\
  0.3   & -1.0435  & -0.2604  & -0.0639  & -0.0157  & -0.0039
        & -0.0010  & -0.0002  &  0.0000  & 0 \\
  0.4   & -0.3771  & -0.0521  & -0.0072  & -0.0010  & -0.0001
        &  0.0000  &  0.0000  &  0.0000  & 0 \\
  0.5   & -0.1165  & -0.0084  & -0.0006  &  0.0000  &  0.0000
        &  0.0000  &  0.0000  &  0.0000  & 0 \\
  0.6   & -0.0313  & -0.0011  &  0.0000  &  0.0000  &  0.0000
        &  0.0000  &  0.0000  &  0.0000  & 0 \\
  0.7   & -0.0074  & -0.0001  &  0.0000  &  0.0000  &  0.0000
        &  0.0000  &  0.0000  &  0.0000  & 0 \\
  0.8   & -0.0017  &  0.0000  &  0.0000  &  0.0000  &  0.0000
        &  0.0000  &  0.0000  &  0.0000  & 0 \\
  0.9   & -0.0004  &  0.0000  &  0.0000  &  0.0000  &  0.0000
        &  0.0000  &  0.0000  &  0.0000  & 0 \\
  1.0   &  0.0000  &  0.0000  &  0.0000  &  0.0000  &  0.0000
        &  0.0000  &  0.0000  &  0.0000  & 0 \\
  1.1   &  0.0002  &  0.0000  &  0.0000  &  0.0000  &  0.0000
        &  0.0000  &  0.0000  &  0.0000  & 0 \\
  1.2   &  0.0003  &  0.0000  &  0.0000  &  0.0000  &  0.0000
        &  0.0000  &  0.0000  &  0.0000  & 0 \\
  1.3   &  0.0005  &  0.0000  &  0.0000  &  0.0000  &  0.0000
        &  0.0000  &  0.0000  &  0.0000  & 0 \\
  1.4   &  0.0011  &  0.0000  &  0.0000  &  0.0000  &  0.0000
        &  0.0000  &  0.0000  &  0.0000  & 0 \\
  1.5   &  0.0030  & -0.0002  &  0.0000  &  0.0000  &  0.0000
        &  0.0000  &  0.0000  &  0.0000  & 0 \\
  1.6   &  0.0115  & -0.0031  & -0.0007  & -0.0001  &  0.0000
        &  0.0000  &  0.0000  &  0.0000  & 0 \\
  1.7   &  0.0279  & -0.0171  & -0.0082  & -0.0026  & -0.0007
        & -0.0002  &  0.0000  &  0.0000  & 0 \\
  1.8   & -0.0434  & -0.0836  & -0.0583  & -0.0325  & -0.0161
        & -0.0074  & -0.0033  & -0.0001  & 0 \\
  1.9   & -0.8296  & -0.6069  & -0.4382  & -0.3152  & -0.2253
        & -0.1599  & -0.1125  & -0.0256  & 0 \\ \hline
\end{tabular}
\end{ruledtabular}
\end{table*}

\section{Conclusions}
\label{sec:concl}

In this article we have presented the calculations of a few one-loop amplitudes
for domain-wall fermions at finite $N_s$ using improved gauge actions, with the
intention of studying in perturbation theory the phenomenon of reduction of 
chiral violations associated with these gauge actions. In particular, we have 
considered three quantities whose differences from their (vanishing) values at 
$N_s=\infty$ can provide some significant estimates of chiral violations: 
the residual mass, the difference $\Delta$ between the vector and axial-vector 
renormalization constants, and $c_{mix}$, a power-divergent mixing of a 
deep-inelastic operator which is entirely due to the breaking of chirality. 

It is then useful to compare the results presented in this paper with the ones
computed in Ref. \cite{Capitani:2006kw} for the simple plaquette action, and 
see whether and how much they have decreased when the coupling is kept 
constant. Our calculations show that also 
in the framework of one-loop perturbation theory the use of improved gauge 
actions can indeed suppress the residual mass. The largest suppressions are 
produced by the DBW2 action, and this dramatic decrease, of about one order 
of magnitude, is due to the combined effects of the changes in the (order-zero)
tadpole and half-circle diagrams of the self-energy. Thus, our perturbative 
results, when used at the same values of $g_0$ for the various actions, 
may qualitatively confirm what is known from Monte Carlo simulations about 
the consequences of using improved gauge actions. 

On the other hand, if one compares the various actions at the same energy 
scale, perturbation theory gives rather puzzling results (especially in the 
case of the DBW2 action), and its use appears to be problematic (at least 
at the one-loop level), also because of the large values that the gauge 
coupling assumes at the scale of 2 GeV.

We have also found that the effects of improved gauge actions can also be
different depending on the quantity studied. When $g_0$ is kept fixed, there is
indeed a dramatic suppression of the residual mass and of the power-divergent 
mixing coefficient $c_{mix}$ for the DBW2 action, whereas for $\Delta$ only 
very small changes are observed when instead of the plaquette any of the 
improved gauge actions is used. 
A lesson from this could be that some (but by no means not all) 
quantities which measure chiral violations but are of higher order in 
$m_{res}$ will get little or no reduction even with improved gauge actions.

\begin{acknowledgments}
I am grateful for the support by Fonds zur F\"orderung der Wissenschaftlichen 
Forschung in \"Osterreich (FWF), Project P16310-N08. I also thank Rainer 
Sommer for reminding me to compare the various actions at their proper 
values of $\beta$.
\end{acknowledgments}

\appendix*

\section{Some plaquette results}

In this Appendix we collect some useful reference results derived in 
\cite{Capitani:2006kw} for the plaquette action. Values of the tree-level 
residual mass (valid for any kind of gluon action) are reported in Table 
\ref{tab:residual_tree}. The plaquette numbers for $\Sigma_0$ in Feynman gauge 
are shown in Table \ref{tab:sigma0feynman}, and the corresponding one-loop 
residual mass for $\beta = 5.2$ and $\beta = 6.0$ in Landau gauge are reported 
in Tables \ref{tab:residual2l} and \ref{tab:residual1l}. 

The plaquette numbers for $\Delta$ and $c_{mix}$ in Feynman gauge are finally 
shown in Tables \ref{tab:va} and \ref{tab:d1mix} respectively.

\begin{table*}[htp]
\caption{\label{tab:residual_tree}
Residual mass at tree level in lattice units (multiplied for $16 \pi^2$).}
\begin{ruledtabular}    
\begin{tabular}{|c|rrrrrrrrr|} 
$M$ & $N_s=8$ & $N_s=12$ & $N_s=16$ & $N_s=20$ & $N_s=24$ & $N_s=28$ 
    & $N_s=32$ & $N_s=48$ & $N_s=\infty$  
      \vspace{0.05cm} \\ \hline \vspace{-0.3cm} \\  
  0.1   &-12.91556 & -8.47390 & -5.55973 & -3.64774 & -2.39328 
        & -1.57023 & -1.03023 & -0.19090 &  0  \\
  0.2   & -9.53767 & -3.90663 & -1.60015 & -0.65542 & -0.26846      
        & -0.10996 & -0.04504 & -0.00127 &  0  \\
  0.3   & -4.64274 & -1.11472 & -0.26764 & -0.06426 & -0.01543 
        & -0.00370 & -0.00089 &  0.00000 &  0  \\
  0.4   & -1.69750 & -0.22000 & -0.02851 & -0.00370 & -0.00048 
        & -0.00006 & -0.00001 &  0.00000 &  0  \\
  0.5   & -0.46264 & -0.02891 & -0.00181 & -0.00011 & -0.00001 
        &  0.00000 &  0.00000 &  0.00000 &  0  \\
  0.6   & -0.08693 & -0.00223 & -0.00006 &  0.00000 &  0.00000 
        &  0.00000 &  0.00000 &  0.00000 &  0  \\
  0.7   & -0.00943 & -0.00008 &  0.00000 &  0.00000 &  0.00000 
        &  0.00000 &  0.00000 &  0.00000 &  0  \\
  0.8   & -0.00039 &  0.00000 &  0.00000 &  0.00000 &  0.00000 
        &  0.00000 &  0.00000 &  0.00000 &  0  \\
  0.9   &  0.00000 &  0.00000 &  0.00000 &  0.00000 &  0.00000 
        &  0.00000 &  0.00000 &  0.00000 &  0  \\
  1.0   &  0.00000 &  0.00000 &  0.00000 &  0.00000 &  0.00000 
        &  0.00000 &  0.00000 &  0.00000 &  0  \\
  1.1   &  0.00000 &  0.00000 &  0.00000 &  0.00000 &  0.00000 
        &  0.00000 &  0.00000 &  0.00000 &  0  \\
  1.2   & -0.00039 &  0.00000 &  0.00000 &  0.00000 &  0.00000 
        &  0.00000 &  0.00000 &  0.00000 &  0  \\
  1.3   & -0.00943 & -0.00008 &  0.00000 &  0.00000 &  0.00000 
        &  0.00000 &  0.00000 &  0.00000 &  0  \\
  1.4   & -0.08693 & -0.00223 & -0.00006 &  0.00000 &  0.00000 
        &  0.00000 &  0.00000 &  0.00000 &  0  \\
  1.5   & -0.46264 & -0.02891 & -0.00181 & -0.00011 & -0.00001 
        &  0.00000 &  0.00000 &  0.00000 &  0  \\
  1.6   & -1.69750 & -0.22000 & -0.02851 & -0.00370 & -0.00048 
        & -0.00006 & -0.00001 &  0.00000 &  0  \\
  1.7   & -4.64274 & -1.11472 & -0.26764 & -0.06426 & -0.01543 
        & -0.00370 & -0.00089 &  0.00000 &  0  \\
  1.8   & -9.53767 & -3.90663 & -1.60015 & -0.65542 & -0.26846      
        & -0.10996 & -0.04504 & -0.00127 &  0  \\
  1.9   &-12.91556 & -8.47390 & -5.55973 & -3.64774 & -2.39328 
        & -1.57023 & -1.03023 & -0.19090 &  0  \\ \hline    
\end{tabular}
\end{ruledtabular}
\end{table*}

\begin{table*}[htp]
\caption{\label{tab:sigma0feynman}
Coefficient of $\bar g^2$ for the complete result of $\Sigma_0$, in Feynman 
gauge for the plaquette gauge action.}
\begin{ruledtabular}    
\begin{tabular}{|c|rrrrrrrrr|} 
$M$ & $N_s=8$ & $N_s=12$ & $N_s=16$ & $N_s=20$ & $N_s=24$ & $N_s=28$ 
    & $N_s=32$ & $N_s=48$ & $N_s=\infty$  
      \vspace{0.05cm} \\ \hline \vspace{-0.3cm} \\  
  0.1   & 17.85919  & 19.67880  & 19.25277  & 17.32109  & 14.66546
        & 11.87550  &  9.30218  &  2.86309  & 0  \\
  0.2   & 21.82264  & 15.20660  &  8.97397  &  4.82019  &  2.44504
        &  1.19446  &  0.56831  &  0.02488  & 0  \\
  0.3   & 15.28940  &  5.98758  &  1.99997  &  0.61538  &  0.18022
        &  0.05106  &  0.01412  &  0.00007  & 0  \\
  0.4   &  7.55847  &  1.55410  &  0.27613  &  0.04548  &  0.00713
        &  0.00108  &  0.00016  &  0.00000  & 0  \\
  0.5   &  2.79190  &  0.27543  &  0.02358  &  0.00186  &  0.00014
        &  0.00001  &  0.00000  &  0.00000  & 0  \\
  0.6   &  0.76252  &  0.03191  &  0.00115  &  0.00004  &  0.00000
        &  0.00000  &  0.00000  &  0.00000  & 0  \\
  0.7   &  0.15015  &  0.00250  &  0.00004  &  0.00000  &  0.00000
        &  0.00000  &  0.00000  &  0.00000  & 0  \\
  0.8   &  0.02561  &  0.00031  &  0.00001  &  0.00000  &  0.00000
        &  0.00000  &  0.00000  &  0.00000  & 0  \\
  0.9   &  0.00774  &  0.00012  &  0.00000  &  0.00000  &  0.00000
        &  0.00000  &  0.00000  &  0.00000  & 0  \\
  1.0   &  0.00418  &  0.00008  &  0.00000  &  0.00000  &  0.00000
        &  0.00000  &  0.00000  &  0.00000  & 0  \\
  1.1   &  0.00283  &  0.00007  &  0.00000  &  0.00000  &  0.00000
        &  0.00000  &  0.00000  &  0.00000  & 0  \\
  1.2   & -0.00172  &  0.00006  &  0.00000  &  0.00000  &  0.00000
        &  0.00000  &  0.00000  &  0.00000  & 0  \\
  1.3   & -0.06252  & -0.00073  &  0.00000  &  0.00000  &  0.00000
        &  0.00000  &  0.00000  &  0.00000  & 0  \\
  1.4   & -0.44327  & -0.01733  & -0.00060  & -0.00002  &  0.00000
        &  0.00000  &  0.00000  &  0.00000  & 0  \\
  1.5   & -1.85863  & -0.18051  & -0.01531  & -0.00122  & -0.00009
        & -0.00001  &  0.00000  &  0.00000  & 0  \\
  1.6   & -5.42629  & -1.12306  & -0.20001  & -0.03298  & -0.00521
        & -0.00080  & -0.00012  &  0.00000  & 0  \\
  1.7   &-11.49207  & -4.61264  & -1.55867  & -0.48269  & -0.14194
        & -0.04034  & -0.01120  & -0.00006  & 0  \\
  1.8   &-16.77186  &-12.20090  & -7.36730  & -4.00994  & -2.05105
        & -1.00765  & -0.48138  & -0.02129  & 0  \\
  1.9   &-13.69620  &-15.97157  &-16.18935  &-14.90821  &-12.82561
        &-10.50396  & -8.29629  & -2.59956  & 0  \\
 \hline    
\end{tabular}
\end{ruledtabular}
\end{table*}

\begin{table*}[htp]
\caption{\label{tab:residual2l}
Residual mass in lattice units at $\beta=5.2$, in Landau gauge
for the plaquette gauge action.} 
\begin{ruledtabular}    
\begin{tabular}{|c|rrrrrrrrr|} 
$M$ & $N_s=8$ & $N_s=12$ & $N_s=16$ & $N_s=20$ & $N_s=24$ & $N_s=28$ 
    & $N_s=32$ & $N_s=48$ & $N_s=\infty$  
      \vspace{0.05cm} \\ \hline \vspace{-0.3cm} \\  
  0.1   & -0.25014  & -0.24077  & -0.21923  & -0.18923  & -0.15614
        & -0.12430  & -0.09621  & -0.02889  & 0  \\
  0.2   & -0.26738  & -0.17001  & -0.09619  & -0.05047  & -0.02523
        & -0.01220  & -0.00576  & -0.00025  & 0  \\
  0.3   & -0.17502  & -0.06438  & -0.02089  & -0.00632  & -0.00183
        & -0.00051  & -0.00014  &  0.00000  & 0  \\
  0.4   & -0.08294  & -0.01629  & -0.00283  & -0.00046  & -0.00007
        & -0.00001  &  0.00000  &  0.00000  & 0  \\
  0.5   & -0.02966  & -0.00283  & -0.00024  & -0.00002  &  0.00000
        &  0.00000  &  0.00000  &  0.00000  & 0  \\
  0.6   & -0.00787  & -0.00032  & -0.00001  &  0.00000  &  0.00000
        &  0.00000  &  0.00000  &  0.00000  & 0  \\
  0.7   & -0.00151  & -0.00002  &  0.00000  &  0.00000  &  0.00000
        &  0.00000  &  0.00000  &  0.00000  & 0  \\
  0.8   & -0.00025  &  0.00000  &  0.00000  &  0.00000  &  0.00000
        &  0.00000  &  0.00000  &  0.00000  & 0  \\
  0.9   & -0.00008  &  0.00000  &  0.00000  &  0.00000  &  0.00000
        &  0.00000  &  0.00000  &  0.00000  & 0  \\
  1.0   & -0.00004  &  0.00000  &  0.00000  &  0.00000  &  0.00000
        &  0.00000  &  0.00000  &  0.00000  & 0  \\
  1.1   & -0.00003  &  0.00000  &  0.00000  &  0.00000  &  0.00000
        &  0.00000  &  0.00000  &  0.00000  & 0  \\
  1.2   &  0.00002  &  0.00000  &  0.00000  &  0.00000  &  0.00000
        &  0.00000  &  0.00000  &  0.00000  & 0  \\
  1.3   &  0.00056  &  0.00001  &  0.00000  &  0.00000  &  0.00000
        &  0.00000  &  0.00000  &  0.00000  & 0  \\
  1.4   &  0.00387  &  0.00016  &  0.00001  &  0.00000  &  0.00000
        &  0.00000  &  0.00000  &  0.00000  & 0  \\
  1.5   &  0.01565  &  0.00161  &  0.00014  &  0.00001  &  0.00000
        &  0.00000  &  0.00000  &  0.00000  & 0  \\
  1.6   &  0.04357  &  0.00979  &  0.00181  &  0.00030  &  0.00005
        &  0.00001  &  0.00000  &  0.00000  & 0  \\
  1.7   &  0.08592  &  0.03889  &  0.01378  &  0.00438  &  0.00131
        &  0.00038  &  0.00011  &  0.00000  & 0  \\
  1.8   &  0.10875  &  0.09702  &  0.06302  &  0.03555  &  0.01857
        &  0.00925  &  0.00446  &  0.00020  & 0  \\
  1.9   &  0.05763  &  0.10669  &  0.12612  &  0.12479  &  0.11169
        &  0.09373  &  0.07524  &  0.02433  & 0  \\ \hline    
\end{tabular}
\end{ruledtabular}
\end{table*}

\begin{table*}[htp]
\caption{\label{tab:residual1l}
Residual mass in lattice units at $\beta=6$, in Landau gauge
for the plaquette gauge action.} 
\begin{ruledtabular}    
\begin{tabular}{|c|rrrrrrrrr|} 
$M$ & $N_s=8$ & $N_s=12$ & $N_s=16$ & $N_s=20$ & $N_s=24$ & $N_s=28$ 
    & $N_s=32$ & $N_s=48$ & $N_s=\infty$  
      \vspace{0.05cm} \\ \hline \vspace{-0.3cm} \\  
  0.1   & -0.22770  & -0.21582  & -0.19469  & -0.16708  & -0.13735
        & -0.10905  & -0.08425  & -0.02520  & 0  \\
  0.2   & -0.23978  & -0.15064  & -0.08471  & -0.04430  & -0.02209
        & -0.01067  & -0.00503  & -0.00022  & 0  \\
  0.3   & -0.15560  & -0.05674  & -0.01833  & -0.00553  & -0.00160
        & -0.00045  & -0.00012  &  0.00000  & 0  \\
  0.4   & -0.07331  & -0.01430  & -0.00248  & -0.00040  & -0.00006
        & -0.00001  &  0.00000  &  0.00000  & 0  \\
  0.5   & -0.02610  & -0.00247  & -0.00021  & -0.00002  &  0.00000
        &  0.00000  &  0.00000  &  0.00000  & 0  \\
  0.6   & -0.00690  & -0.00028  & -0.00001  &  0.00000  &  0.00000
        &  0.00000  &  0.00000  &  0.00000  & 0  \\
  0.7   & -0.00132  & -0.00002  &  0.00000  &  0.00000  &  0.00000
        &  0.00000  &  0.00000  &  0.00000  & 0  \\
  0.8   & -0.00022  &  0.00000  &  0.00000  &  0.00000  &  0.00000
        &  0.00000  &  0.00000  &  0.00000  & 0  \\
  0.9   & -0.00007  &  0.00000  &  0.00000  &  0.00000  &  0.00000
        &  0.00000  &  0.00000  &  0.00000  & 0  \\
  1.0   & -0.00004  &  0.00000  &  0.00000  &  0.00000  &  0.00000
        &  0.00000  &  0.00000  &  0.00000  & 0  \\
  1.1   & -0.00002  &  0.00000  &  0.00000  &  0.00000  &  0.00000
        &  0.00000  &  0.00000  &  0.00000  & 0  \\
  1.2   &  0.00001  &  0.00000  &  0.00000  &  0.00000  &  0.00000
        &  0.00000  &  0.00000  &  0.00000  & 0  \\
  1.3   &  0.00048  &  0.00001  &  0.00000  &  0.00000  &  0.00000
        &  0.00000  &  0.00000  &  0.00000  & 0  \\
  1.4   &  0.00328  &  0.00014  &  0.00000  &  0.00000  &  0.00000
        &  0.00000  &  0.00000  &  0.00000  & 0  \\
  1.5   &  0.01317  &  0.00137  &  0.00012  &  0.00001  &  0.00000
        &  0.00000  &  0.00000  &  0.00000  & 0  \\
  1.6   &  0.03632  &  0.00830  &  0.00154  &  0.00026  &  0.00004
        &  0.00001  &  0.00000  &  0.00000  & 0  \\
  1.7   &  0.07054  &  0.03276  &  0.01172  &  0.00374  &  0.00112
        &  0.00032  &  0.00009  &  0.00000  & 0  \\
  1.8   &  0.08620  &  0.08078  &  0.05326  &  0.03026  &  0.01587
        &  0.00793  &  0.00383  &  0.00017  & 0  \\
  1.9   &  0.03905  &  0.08531  &  0.10461  &  0.10507  &  0.09478
        &  0.07991  &  0.06434  &  0.02093  & 0  \\ \hline    
\end{tabular}
\end{ruledtabular}
\end{table*}

\begin{table*}[htp]
\caption{\label{tab:va}
Coefficient of $\bar g^2$ for $\Delta$, in Feynman gauge for the plaquette 
gauge action.}
\begin{ruledtabular}    
\begin{tabular}{|c|rrrrrrrrr|} 
$M$ & $N_s=8$ & $N_s=12$ & $N_s=16$ & $N_s=20$ & $N_s=24$ & $N_s=28$ 
    & $N_s=32$ & $N_s=48$ & $N_s=\infty$  
      \vspace{0.05cm} \\ \hline \vspace{-0.3cm} \\  
  0.1   &  6.9399  &  4.4434  &  2.6941  &  1.5749  &  0.8916  
        &  0.4899  &  0.2622  &  0.0179  & 0 \\
  0.2   &  2.3272  &  0.7211  &  0.1945  &  0.0477  &  0.0110 
        &  0.0025  &  0.0005  &  0.0000  & 0 \\
  0.3   &  0.5894  &  0.0668  &  0.0064  &  0.0006  &  0.0000  
        &  0.0000  &  0.0000  &  0.0000  & 0 \\
  0.4   &  0.0994  &  0.0034  &  0.0001  &  0.0000  &  0.0000
        &  0.0000  &  0.0000  &  0.0000  & 0 \\
  0.5   &  0.0106  &  0.0001  &  0.0000  &  0.0000  &  0.0000
        &  0.0000  &  0.0000  &  0.0000  & 0 \\
  0.6   &  0.0007  &  0.0000  &  0.0000  &  0.0000  &  0.0000
        &  0.0000  &  0.0000  &  0.0000  & 0 \\
  0.7   &  0.0000  &  0.0000  &  0.0000  &  0.0000  &  0.0000
        &  0.0000  &  0.0000  &  0.0000  & 0 \\
  0.8   &  0.0000  &  0.0000  &  0.0000  &  0.0000  &  0.0000
        &  0.0000  &  0.0000  &  0.0000  & 0 \\
  0.9   &  0.0000  &  0.0000  &  0.0000  &  0.0000  &  0.0000
        &  0.0000  &  0.0000  &  0.0000  & 0 \\
  1.0   &  0.0000  &  0.0000  &  0.0000  &  0.0000  &  0.0000
        &  0.0000  &  0.0000  &  0.0000  & 0 \\
  1.1   &  0.0000  &  0.0000  &  0.0000  &  0.0000  &  0.0000
        &  0.0000  &  0.0000  &  0.0000  & 0 \\
  1.2   &  0.0000  &  0.0000  &  0.0000  &  0.0000  &  0.0000
        &  0.0000  &  0.0000  &  0.0000  & 0 \\
  1.3   &  0.0000  &  0.0000  &  0.0000  &  0.0000  &  0.0000
        &  0.0000  &  0.0000  &  0.0000  & 0 \\
  1.4   &  0.0005  &  0.0000  &  0.0000  &  0.0000  &  0.0000
        &  0.0000  &  0.0000  &  0.0000  & 0 \\
  1.5   &  0.0083  &  0.0001  &  0.0000  &  0.0000  &  0.0000
        &  0.0000  &  0.0000  &  0.0000  & 0 \\
  1.6   &  0.0701  &  0.0029  &  0.0001  &  0.0000  &  0.0000
        &  0.0000  &  0.0000  &  0.0000  & 0 \\
  1.7   &  0.3140  &  0.0501  &  0.0056  &  0.0005  &  0.0000
        &  0.0000  &  0.0000  &  0.0000  & 0 \\
  1.8   &  0.8354  &  0.3672  &  0.1288  &  0.0374  &  0.0096
        &  0.0022  &  0.0005  &  0.0000  & 0 \\
  1.9   &  2.2075  &  1.5323  &  1.0348  &  0.6904  &  0.4490
        &  0.2811  &  0.1686  &  0.0153  & 0 \\ \hline
\end{tabular}
\end{ruledtabular}
\end{table*}

\begin{table*}[htp]
\caption{\label{tab:d1mix}
Coefficient of $\bar g^2$ for $c_{mix}$, in Feynman gauge for the plaquette 
gauge action.}
\begin{ruledtabular}    
\begin{tabular}{|c|rrrrrrrrr|} 
$M$ & $N_s=8$ & $N_s=12$ & $N_s=16$ & $N_s=20$ & $N_s=24$ & $N_s=28$ 
    & $N_s=32$ & $N_s=48$ & $N_s=\infty$  
      \vspace{0.05cm} \\ \hline \vspace{-0.3cm} \\  
  0.1   &-14.1244  & -8.4448  & -5.3682  & -3.4884  & -2.2846
        & -1.4996  & -0.9848  & -0.1829  & 0 \\
  0.2   & -5.8166  & -2.4014  & -0.9910  & -0.4075  & -0.1674
        & -0.0688  & -0.0283  & -0.0008  & 0 \\
  0.3   & -2.4733  & -0.6070  & -0.1475  & -0.0359  & -0.0087
        & -0.0021  & -0.0005  &  0.0000  & 0 \\
  0.4   & -0.9361  & -0.1265  & -0.0169  & -0.0022  & -0.0003
        &  0.0000  &  0.0000  &  0.0000  & 0 \\
  0.5   & -0.3094  & -0.0213  & -0.0014  & -0.0001  &  0.0000
        &  0.0000  &  0.0000  &  0.0000  & 0 \\
  0.6   & -0.0893  & -0.0029  & -0.0001  &  0.0000  &  0.0000
        &  0.0000  &  0.0000  &  0.0000  & 0 \\
  0.7   & -0.0231  & -0.0004  &  0.0000  &  0.0000  &  0.0000
        &  0.0000  &  0.0000  &  0.0000  & 0 \\
  0.8   & -0.0059  & -0.0001  &  0.0000  &  0.0000  &  0.0000
        &  0.0000  &  0.0000  &  0.0000  & 0 \\
  0.9   & -0.0015  &  0.0000  &  0.0000  &  0.0000  &  0.0000
        &  0.0000  &  0.0000  &  0.0000  & 0 \\
  1.0   &  0.0000  &  0.0000  &  0.0000  &  0.0000  &  0.0000
        &  0.0000  &  0.0000  &  0.0000  & 0 \\
  1.1   &  0.0008  &  0.0000  &  0.0000  &  0.0000  &  0.0000
        &  0.0000  &  0.0000  &  0.0000  & 0 \\
  1.2   &  0.0014  &  0.0000  &  0.0000  &  0.0000  &  0.0000
        &  0.0000  &  0.0000  &  0.0000  & 0 \\
  1.3   &  0.0031  &  0.0001  &  0.0000  &  0.0000  &  0.0000
        &  0.0000  &  0.0000  &  0.0000  & 0 \\
  1.4   &  0.0127  &  0.0003  &  0.0000  &  0.0000  &  0.0000  
        &  0.0000  &  0.0000  &  0.0000  & 0 \\
  1.5   &  0.0546  &  0.0029  &  0.0002  &  0.0000  &  0.0000
        &  0.0000  &  0.0000  &  0.0000  & 0 \\
  1.6   &  0.1969  &  0.0207  &  0.0024  &  0.0003  &  0.0000
        &  0.0000  &  0.0000  &  0.0000  & 0 \\
  1.7   &  0.5774  &  0.1129  &  0.0229  &  0.0049  &  0.0011
        &  0.0002  &  0.0001  &  0.0000  & 0 \\
  1.8   &  1.3844  &  0.4817  &  0.1715  &  0.0615  &  0.0224
        &  0.0083  &  0.0031  &  0.0001  & 0 \\
  1.9   &  2.9825  &  1.5917  &  0.9340  &  0.5665  &  0.3480
        &  0.2148  &  0.1328  &  0.0198  & 0 \\ \hline
\end{tabular}
\end{ruledtabular}
\end{table*}

\end{document}